\documentclass[onecolumn,journal]{IEEEtran}
\usepackage[a4paper, total={6.7in, 9.5in}]{geometry}
\usepackage{amsmath,amsfonts}
\usepackage{algorithmic}
\usepackage{algorithm}
\usepackage{array}
\usepackage[caption=false,font=normalsize,labelfont=sf,textfont=sf]{subfig}
\usepackage{textcomp}
\usepackage{stfloats}
\usepackage{url}
\usepackage{verbatim}
\usepackage{graphicx}
\usepackage{cite}
\usepackage{amsmath,amssymb,amsfonts,mathtools}
\usepackage{algorithmic}
\usepackage{graphicx}
\usepackage{textcomp}
\usepackage{tikz}
\usepackage{siunitx}
\usepackage{arydshln}
\usepackage{balance}
\usepackage{lipsum}
\usepackage{arydshln}
\usepackage{booktabs}
\usepackage{bbold}
\usepackage{multirow}
\usepackage{marginnote}
\usepackage{url}
\usepackage{color, soul}
\usepackage{lipsum}
\usepackage{multicol}
\usepackage[makeroom]{cancel}
\usepackage{subfig}

\newtheorem{ass}{Assumption}[section]
\newtheorem{remark}{Remark}[section]

\usepackage{color,soul}

\usepackage{amsmath,bm}
\usepackage{tabularx}
\usepackage{xcolor}
\usepackage{eurosym}
\usepackage{mathtools}

\begin{document}
\title{Physics-informed Neural Network Modelling and Predictive Control of District Heating Systems}
\author{Laura Boca de Giuli, Alessio La Bella, and Riccardo Scattolini
	\thanks{Laura Boca de Giuli, Alessio La Bella and Riccardo Scattolini are with the Dipartimento di Elettronica, Informazione e Bioingegneria,
	Politecnico di Milano, 20133 Milan, Italy (e-mails: \textsl{laura.bocadegiuli@polimi.it}, \textsl{alessio.labella@polimi.it},  and
	\textsl{riccardo.scattolini@polimi.it}).}}



\maketitle

\begin{abstract}
This paper addresses the data-based modelling and optimal control of District Heating Systems (DHSs). 
Physical models of such large-scale networked systems are governed by complex nonlinear equations that require a large amount of parameters, leading to potential computational issues in optimizing their operation.
A novel methodology is hence proposed, exploiting operational data and available physical knowledge to attain accurate and computationally efficient DHSs dynamic models. The proposed idea consists in leveraging multiple Recurrent Neural Networks (RNNs) and in embedding the physical topology of the DHS network in their interconnections. With respect to standard RNN approaches, the resulting modelling methodology, denoted as Physics-Informed RNN (PI-RNN), enables to achieve faster training procedures and higher modelling accuracy, even when reduced-dimension models are exploited. The developed PI-RNN modelling technique paves the way for the design of a Nonlinear Model Predictive Control (NMPC) regulation strategy, enabling, with limited computational time, to minimize production costs, to increase system efficiency and to respect operative constraints over the whole DHS network. The proposed methods are tested in simulation on a DHS benchmark referenced in the literature, showing promising results from the modelling and control perspective.
\end{abstract}


\section{Introduction}
The growing issue of climate change calls for cutting-edge solutions to substantially reduce carbon emissions. In this context, District Heating Systems (DHSs), given their high efficiency, are recognized as crucial to reach the energy transition objectives. In fact, the European Commission considers this technology necessary to meet the 2050 decarbonization targets \cite{EU_DHS}, with the aim of covering at least 50\% of the heating demand in most European countries \cite{paardekooper2018heat}. A DHS is generally composed of a heating station, comprising different thermal generators, and of an insulated water pipeline network, transferring the generated heat to thermal loads (e.g., residential and commercial users), which, exploiting local heat exchangers, absorb the delivered heat and use it for indoor heating and domestic hot water. DHSs are typically operated by heuristic rule-based control strategies, which however do not exploit their full efficiency potential, implying the necessity to design advanced optimization-based control strategies \cite{buffa2021advanced}. This is not a trivial task though, as DHSs are  large-scale systems governed by complex nonlinear dynamical equations (e.g., describing transport phenomena in thermo-hydraulic networks), entailing a significant effort to compute their optimal operation and to develop accurate physical models, also due to the considerable number of necessary parameters (e.g., pipes lengths, diameters, friction coefficients, etc.) \cite{krug2021nonlinear, bastida2021modelling}. 

To overcome these issues, it is here proposed to rely on identification methods with the purpose of obtaining computationally-efficient and accurate models from operational data, which are typically widely available in DHSs. More specifically, Recurrent Neural Networks (RNNs) are employed, being particularly suited to model nonlinear dynamical systems \cite{bemporad2022recurrent, bonassi2022recurrent}. It is worth remarking that RNNs generally do not exploit any physical insight on the identified system: this may lead to the need of large datasets, time-consuming training procedures, or even unreliable data-based models. On the other hand, besides operational data, in engineering systems there is usually the availability of some physical knowledge, which is worth being used to develop physically consistent data-based models. This has motivated the design of a novel Physics-Informed RNN (PI-RNN) modelling methodology for DHSs, enabling to achieve enhanced identification performances and efficient training procedures. In particular, the commonly known information about the physical topology of the DHS network (i.e., how thermal loads and generators are interconnected) is exploited to develop a PI-RNN model with an analogous topological structure. It is also shown that the developed PI-RNN model can be effectively employed to design a Nonlinear Model Predictive Control (NMPC) regulator, enabling to minimize production costs and to increase the system efficiency while respecting the desired operational constraints (e.g., temperature limits over the network). 

\subsection{Related work}
The detailed modelling of DHSs is addressed in \cite{machado2022modeling}, focusing on the stability of their nonlinear dynamics. The physical modelling and optimal operation of DHSs is discussed in \cite{krug2021nonlinear}, yet leading to the formulation of a large-scale problem solved using a one-step prediction horizon. In fact, DHS physical models typically include many state variables governed by nonlinear thermo-hydraulic equations, e.g., describing fluid and heat transport in water pipelines, resulting in a modelling complexity hardly tractable by standard optimization-based controllers. To overcome this issue, predictive controllers exploiting simplified models have been proposed in the literature, such as \cite{farahani2016robust,taylor2021model,verrilli2016stochastic}, where the thermal dynamics of the DHS network are not modelled. 
Nevertheless, the accurate modelling of the network thermal dynamics is crucial for the optimal operation of DHS plants for several reasons: \mbox{\textit{(i)} network} temperatures must respect operative constraints due to technical limits of thermal generators and to the proper heat supply to thermal loads (e.g., the water temperature supplied to each load must exceed a minimum lower bound \cite{9817450,krug2021nonlinear}), and \textit{(ii)} network thermal dynamics, if modelled, can be optimized to minimize heat losses and to increase the overall DHS efficiency. 
Other optimization-based control approaches include a dynamical modelling of DHS networks using simplifying assumptions, such as \cite{sandou2005predictive}, where constant transport delays are considered, or \cite{wirtz2021temperature}, where all thermal loads are assumed to be supplied with the same water temperature (i.e., neglecting heat losses over the DHS network). 
Given the huge complexity of detailed physical models for DHS networks and the poor accuracy of simplified ones, data-based methods have been proposed to identify control-oriented and accurate DHS models directly from operational data \cite{JIANG2023121038,la2021data}. In this context, Neural Networks (NNs) have been exploited for modelling and optimally controlling heating and cooling networks, thanks to their enhanced capability of representing nonlinear dynamical systems \cite{terzi2020learning,buffa2020fifth}.
Nevertheless, the mentioned data-based models disregard any available physical insight on the system to be identified, possibly leading to poorly physically consistent and unreliable models. 

Actually, in the scientific community, a growing interest is arising to embed available physical knowledge in NN models, enhancing their physical consistency, accuracy and training procedure \cite{karpatne2022knowledge}. To do that, different approaches have been presented in the literature. For instance, in \cite{jia2019physics, niaki2021physics, mowlavi2023optimal,drgovna2021physics}, the loss function used for the NNs training is modified such that, besides minimizing the prediction error, known physical equations or relationships among variables are induced to be respected. Other methods suggest to incorporate the available physical knowledge directly in the NN architecture \cite{daw2020physics,misyris2021capturing}. In this context, in \cite{bolderman2021physics}, a physics-guided layer, embedding known system dynamics, is placed in parallel to NN hidden layers, improving the modelling performances. Considering the problem of deriving data-based models of interconnected systems, a further method consists in exploiting their physical topology, which is generally known, and interconnecting different NN models accordingly. This idea has been applied to chemical processes in \cite{bonassi2022recurrent,alhajeri2022process}, leveraging the known sequence of operations, and to thermal buildings in \cite{zakwan2022physically}, exploiting the known connections among different thermal zones. This approach is conceptually similar to Graph Neural Networks (GNNs), where different neurons are interconnected by resembling graph-structure data dependencies \cite{scarselli2008graph}. Nevertheless, none of the mentioned physics-informed identification approaches is applied to energy networks and in particular to DHSs, which are commonly characterized by a well-defined topology, and none of them exploits the developed models for the design of computationally efficient and cost-effective NMPC regulators.

\subsection{Main contribution}
In view of the above discussion, a novel PI-RNN modelling methodology for DHSs is proposed, particularly suited for the design of NMPC regulators. The main contributions of the work are hereafter synthesized.
\smallskip
\begin{itemize}
	\item \textit{Physics-informed Neural Network modelling of DHSs:} 
	Given that the DHS network topology is commonly known, this information is leveraged to develop a novel PI-RNN architecture, capable of accurately modelling the main thermal dynamics. More specifically, a different RNN is firstly paired with each section of the DHS network (e.g., with a thermal load and the corresponding supplying pipes), and, subsequently, all RNNs are interconnected resembling the network physical topology. Then, the overall PI-RNN, comprising all the interconnected RNNs, is trained as a unique data-based model, embedding in its architecture the physical dependence among the different DHS network sections. This enables to achieve a faster training procedure and higher modelling accuracy with respect to standard RNN models, even when employing reduced-order PI-RNN models, as witnessed by the numerical results. 
	
	\medskip
	\item \textit{NMPC design for optimal operation of DHSs:} The developed PI-RNN model is exploited for the design of an NMPC regulator, which optimizes the DHS with a prediction horizon of several hours, enabling to minimize production costs, increase system efficiency and comply with operational constraints over the whole DHS, e.g., by providing proper heat delivery to all thermal loads. Moreover, as witnessed by the numerical results, the employed PI-RNN model enables to reduce the NMPC computational complexity not only with respect to physical models but also with respect to standard RNN-based ones.
\end{itemize}
\smallskip
The proposed approach is tested in simulation on a DHS benchmark, i.e., the AROMA DHS \cite{krug2021nonlinear}, showing promising results from the modelling and control perspective.
\\

\subsection{Paper outline}
The paper is organized as follows. A general overview on the DHS physical modelling is presented in Section \ref{sec:dhs}, together with the description of the benchmark case study analysed in this work. Two data-based modelling approaches, i.e., standard RNN and PI-RNN methods, are presented in Section \ref{sec:id}, with a special focus on the proposed physics-informed data-based methodology and its application to the considered DHS benchmark. The formulation of the NMPC regulator exploiting the developed data-based models is described in Section \ref{sec:MPC}. The numerical results regarding the proposed modelling and control methods are reported in Section \ref{sec:results}. Final conclusions are given in Section \ref{sec:concl}.

\subsection{Notation}
Let $\mathbb{R}$  denote the set of real numbers and  $\mathbb{N}$ the one of natural numbers.
Given two vectors of variables ${x,y}\in \mathbb{R}^{n}$, the inequalities between the two, e.g., $x>y$, are intended element-wise, whereas their Kronecker product is indicated with $x \,\otimes\, y$. For a vector $x \in \mathbb{R}^{n} $, its 2-norm is indicated as $\lVert x \rVert_2$, whereas the vectors of corresponding upper and lower bounds are $\overline{x}\in \mathbb{R}^{n}$ and $\underline{x}\in \mathbb{R}^{n}$, respectively, with $\overline{x}>\underline{x}$.
Considering a real variable $a \in \mathbb{R}$, with $a>0$, $b=\lfloor a \rfloor$ is the largest integer less than or equal to $a$, i.e., \mbox{$b \in \mathbb{N} \cup \{0\}$}. Given a sequence of variables $a_1,\hdots,a_n$, and the set of their indexes $\mathcal{N}=\{1,\hdots,n\}$, the vector $a=[a_1,\hdots,a_n]'$ is compactly written as $a={\{a_i\}}_{\forall i \in \mathcal{N}}$. Given a set $\mathcal{N}$, its cardinality is denoted as $n = |\mathcal{N}|$.
Finally, the main physical variables and parameters used in the paper are reported in Table \ref{table:var}. 
\begin{table}[t!]
	\centering
	\caption{Main system variables and parameters.}
	\begin{tabular}{l l}
	\hline \\[-7px]
	\textbf{Symbol} & \textbf{Description} \\\hline\\[-15px] \\
	$c_w$ & Water specific heat coefficient [J/(kg K)]\\[2px]
    $n_c$ & Number of thermal loads \\[2px]
	$P_0$ & Heating station power [W] \\[2px]
	$q_0$ & Heating station water flow [kg/s] \\[2px]
	$T_0^{\,s}$ & Heating station supply temperature [K] \\[2px]
    $T_0^{\,r}$ & Heating station return temperature [K] \\[2px]
    $P_i^{\,c}$ & Load thermal power demand at node $\alpha_i$ [W] \\[2px]	
    $q_i^{\,c}$ & Load water flow at node $\alpha_i$ [kg/s] \\[2px]
    $T_i^{\,c}$ & Load output temperature at node $\alpha_i$ [K] \\[2px]
	$T_i^{c \star}$ & Load reference output temperature at node $\alpha_i$ [K] \\[2px]
	$q_i^{\,s}$ & Supply water flow at node $\alpha_i$ [kg/s] \\[2px]
    $T_i^{\,s}$ & Supply temperature at node $\alpha_i$ [K] \\[2px]
    $q_i^{\,r}$ & Return water flow at node $\alpha_i$ [kg/s] \\[2px]
    $T_i^{\,r}$ & Return temperature at node $\alpha_i$ [K] \\[2px]
    \hline\\[-15px] \\
	\end{tabular}
	\label{table:var}
\end{table}

\section{Problem statement}
\label{sec:dhs}
\subsection{System and main modelling assumptions}
\begin{figure}[t!]
	\centering
	\includegraphics[width=0.6\textwidth]{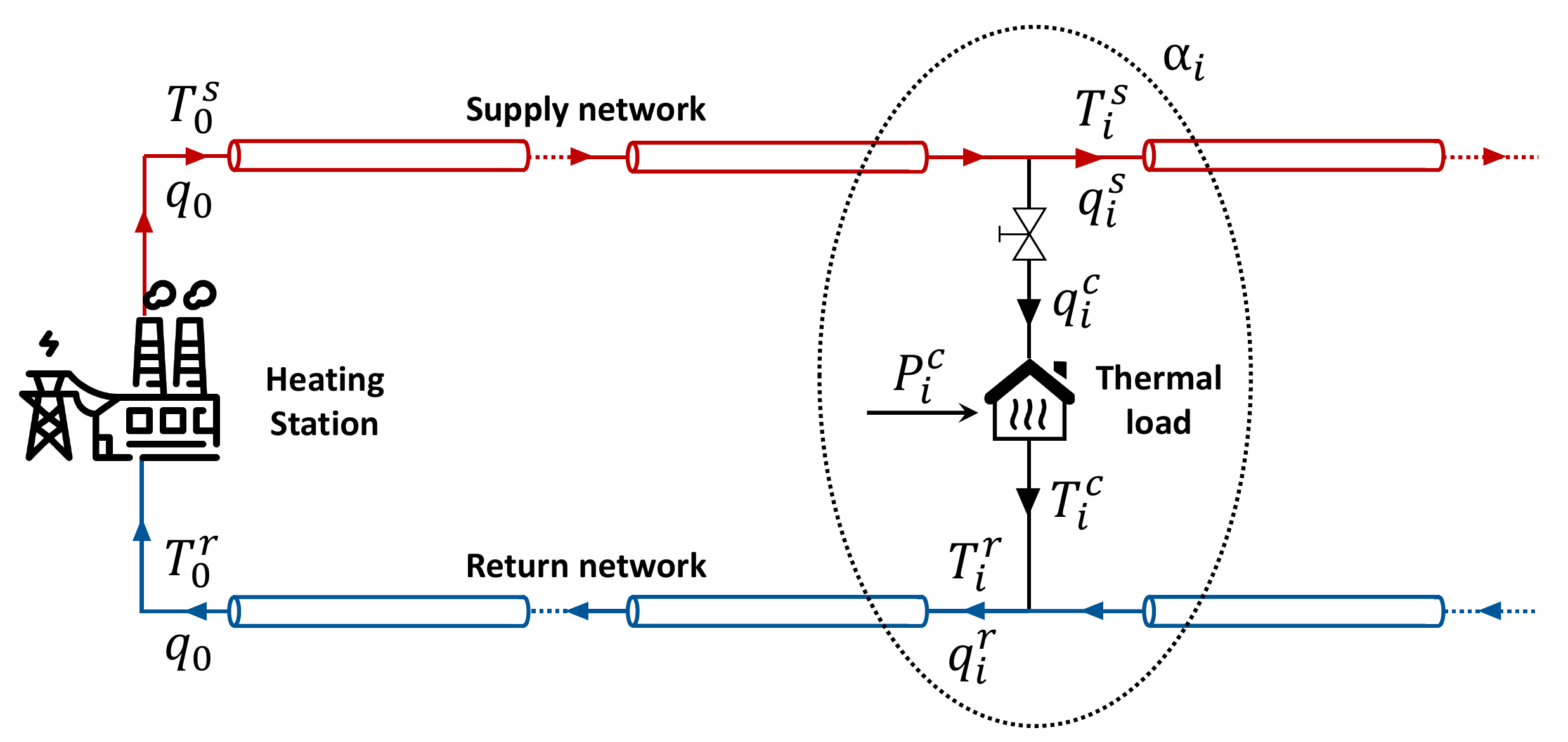}
	\caption{Schematic representation of a district heating system, interconnecting the heating station and the thermal load at node $\alpha_i$.}
	\label{fig:load}
\end{figure}

A DHS typically consists of four main elements, as depicted in Figure \ref{fig:load}: \textit{i)} the supply network, where water at high temperature flows from the heating station to thermal loads, \textit{ii)} the return network, where water at cold temperature flows from thermal loads to the heating station, \textit{iii)} the heating station,  which absorbs water from the return network and inject it at higher temperature into the supply network, and \textit{iv)} the thermal loads (e.g., households or buildings), which absorb water from the supply network, exploiting the delivered heat for internal heating, and inject it into the return network. 

\smallskip
For the sake of clarity, two standard assumptions for DHSs are considered in the following.
\smallskip
\begin{ass}\label{ass:heatingstation}
	The heat generation is centralized, i.e., a single heating station is considered, possibly comprising different thermal generators, \cite{la2023optimal,krug2021nonlinear, verrilli2016stochastic}.
\end{ass}

\smallskip
\begin{ass}\label{ass:topology}
	The supply and return networks have the same physical topology \cite{wang2017hydraulic,machado2022modeling}. 
\end{ass}

\smallskip
Note that the introduced assumptions could be removed in the following at the price of reduced clarity of presentation.

\smallskip 	
Given Assumption \ref{ass:topology}, the DHS can be represented by an oriented graph $\mathcal{G} = (\mathcal{N}, \mathcal{E})$, where $\mathcal{N}$ identifies the set of nodes, whereas $\mathcal{E} \subseteq \mathcal{N} \times \mathcal{N} $ is the set of edges. Each node, denoted as $\alpha_i$ with $i \in \mathcal{N}$, represents a meaningful element of the DHS, e.g., a thermal load or a junction among multiple pipes, and it includes a connection both with the supply and return network, as depicted in Figure \ref{fig:load}. Given Assumption \ref{ass:heatingstation}, without any loss of generality, the node where the heating station is connected is denoted as $\alpha_0$. On the other hand, all the nodes of the DHS network are included in $\mathcal{N}_{\text{net}}$, i.e., $\mathcal{N} = \{0\} \,\cup\, \mathcal{N}_{\text{net}}$, whereas the nodes of the thermal loads connected to the DHS network are included in $\mathcal{N}_c \subseteq \mathcal{N}_{\text{net}}$, with $n_c = |\mathcal{N}_c|$. \\Each edge directed from $\alpha_i$ to $\alpha_j$ is denoted as $e_{ij} = (\alpha_i, \alpha_j)$, with $(i,j) \in\mathcal{E}$. As a convention, each edge is oriented according to the water flow direction in the supply network, assumed to be known, e.g., from available operational data or preprocessing techniques \cite{krug2021nonlinear}. Nevertheless, in case the flow direction in the supply network between $\alpha_i$ and $\alpha_j$ is not fixed, two opposite edges $e_{ij}$ and $e_{ji}$ are defined to interconnect the corresponding nodes.

\smallskip
Since the development of a DHS mathematical model based on physical laws is well known and out of the scope of this paper (the interested reader is referred to \cite{krug2021nonlinear}), the fundamental relations among the main system variables are here briefly presented, as these will be necessary to better describe the proposed identification method.

\smallskip
First of all, as evident from Figure \ref{fig:load}, each load absorbs a water flow $q_i^{\,c}$ from the supply network at temperature $T_{i}^{\,s}$. This water flow goes through an internal heat exchanger absorbing a thermal power $P_i^{\,c}$, then it is injected into the return network at temperature $T_i^{\,c}$, which can be modelled as
\begin{equation}
	\label{eq:Tci}
	\begin{aligned}
		T_i^{\,c}(t)=T_i^{\,s}(t)-\frac{P_i^{\,c}(t)}{q_i^{\,c}(t)\cdot c_w} \;,\; \forall i \in \mathcal{N}_c,
	\end{aligned}
\end{equation}
where $c_w$ is the water specific heat. Note that the thermal load model is static as its dynamical transients are negligible with respect to the DHS network ones.
\\
As discussed in \cite{machado2022decentralized}, the load water flow $q_i^{\,c}$ is supposed to be regulated by a local controller, tracking a constant reference for the load output temperature, indicated as  ${T}_i^{\,c \star}$. This implies that the load water flow can be generally modelled as a function of the load supply and output temperature, the thermal power and the output temperature reference, i.e.,
\begin{equation}
	\label{eq:qci}
	\begin{aligned}
		q_i^{\,c}(t) = \zeta_i^{\,c} (T_i^{\,s}(t),  {T}_i^{\,c}(t), P_i^{\,c}(t), {T}_i^{\,c \star}) \;,\;  \forall i \in \mathcal{N}_c,
	\end{aligned}
\end{equation}
where $\zeta_i^{\,c}$ denotes a generic nonlinear function, whereas $P_i^{\,c}$ and ${T}_i^{\,c \star}$ act as external disturbances.

\smallskip
Before modelling the supply and return network dynamics, the sets of inlet nodes of $\alpha_i$ are defined with respect to the supply and return networks, respectively denoted as $\mathcal{I}_i^{\,s}$ and $\mathcal{I}_i^{\,r}$. More specifically, given that $\mathcal{E}$ is oriented according to the water flow direction of the supply network, it follows that \mbox{$\mathcal{I}_i^{\,s} = \{ j \in \mathcal{N} \, | \, \exists (j, i) \in \mathcal{E} \}$}. On the other hand, in case the water flow directions of the return network are opposite with respect to the supply one, as typical in DHSs \cite{fang2015genetic}, it follows that \mbox{$\mathcal{I}_i^{\,r}=\{ j \in \mathcal{N} \,|\, \exists (i, j) \in \mathcal{E} \} $}. If this does not hold, $\mathcal{I}_i^{\,r}$ can be defined according to the actual water flow directions of the return network.

\smallskip
As shown in Figure \ref{fig:load}, each node of the DHS network is characterized by a net water flow at the supply network, i.e.,  $q_i^{\,s}$, and by one at the return network, i.e, $q_i^{\,r}$. These depend on the water flows at the corresponding inlet network nodes and on the one absorbed, or injected, by the thermal load, if present. Thus, $\forall i \in \mathcal{N}_{\text{net}} $, it holds that
\begin{align}	
	\label{eq:qs}
	q_i^{\,s}(t) = 
	\begin{cases}
		\sum\limits_{\forall j \in I_i^{\,s}(t)}	q_j^{\,s}(t) - q_i^{\,c}(t), \; \text{if} \: i \: \in \mathcal{N}_c \, ,\\
		\sum\limits_{\forall j \in I_i^{\,s}(t)}	q_j^{\,s}(t), \; \text{otherwise}\,,
	\end{cases}
\end{align}

\begin{align}
	\label{eq:qr}
	q_i^{\,r}(t) = 
	\begin{cases}
		\sum\limits_{\forall j \in I_i^{\,r}(t)}	q_j^{\,r}(t) + q_i^{\,c}(t), \; \text{if} \: i \: \in \mathcal{N}_c \, ,\\
		\sum\limits_{\forall j \in I_i^{\,r}(t)}	q_j^{\,r}(t), \; \text{otherwise}.
	\end{cases}
\end{align}

Thermo-hydraulic pipes introduce transport delay effects on temperature profiles over the DHS network. It is possible to describe the physics of water flow in pipelines through 1D Euler equations: in order to write the system in a state-space form, however, the model of each pipe must be discretised both in time and in space (finite volume method), as discussed in details in \cite{krug2021nonlinear}. Hence, each node of the DHS network is characterized by the following temperature dynamics at the supply network, which, $\forall i \in \mathcal{N}_{\text{net}}$, reads as 
\begin{equation}
	\label{eq:Ts}
	\begin{aligned}
		\dot{T}_i^{\,s}(t) = f_i^{\,s} \big (\,z_i^{\,s}(t),\, \{T_j^{\,s}(t), q_j^{\,s}(t)\}_{\forall j \in I_i^{\,s}}\, , T^{\text{ext}}\,\big)\,,
	\end{aligned}
\end{equation}
where $f_i^{\,s}$ is a nonlinear function and $z_i^{\,s}$ is a generic vector employed to represent the internal states of the supply temperature dynamic model, whose definition depends on the pipe spatial discretization \cite{krug2021nonlinear}. Moreover, as evident from \eqref{eq:Ts}, the supply temperature dynamics at each node $\alpha_i$ is influenced by the supply temperatures and water flows of all $\alpha_i$'s inlet nodes. Additionally, $T^{\text{ext}}$ is the external temperature, which corresponds to the ground temperature being DHS pipes typically buried.

Similarly to \eqref{eq:Ts}, each node is characterized by a temperature dynamics at the return network as well, which, $\forall i \in \mathcal{N}_{\text{net}}$, is expressed as
\begin{equation}
	\label{eq:Tr}
	\begin{aligned}
		\dot{T}_i^{\,r}(t) = f_i^{\,r}(z_i^{\,r}(t), T_i^{\,c}(t),\{T_j^{\,r}(t), q_j^{\,r}(t)\}_{\forall j \in I_i^{\,r}}, T^{\text{ext}}),
	\end{aligned}
\end{equation}
where, again, $f_i^{\,r}$ is a nonlinear function, whereas $z_i^{\,r}$ represents the internal states of the return temperature dynamic model. Note that \eqref{eq:Tr}, differently from  \eqref{eq:Ts}, expresses also the dependence on the load output temperature, i.e., $T_i^{\,c}$, as the latter influences the return network dynamics (see Figure \ref{fig:load}).

\smallskip
Finally, regarding the heating station, even though it is typically composed of different thermal generators (boilers, heat pumps, cogenerators, etc.) and storages, here, similarly to \cite{krug2021nonlinear}, its internal configuration is neglected, whereas just its overall power consumption, water flow, return and supply temperature are taken into account. The latter, denoted as $T_0^{\,s}$, does not depend on the DHS network supply nodes, but it is a control variable imposed by the heating station itself \cite{krug2021nonlinear,la2021data}. Moreover, $T_0^{\,s}$ is not here described by a dynamical equation, as in \eqref{eq:Ts}, given that thermal generation is usually characterized by negligible dynamical transients with respect to the DHS network ones \cite{la2023optimal}. On the other hand, the return temperature at the heating station, i.e., $T_0^{\,r}$, is characterized by a dynamical behaviour, which reads as
\begin{equation}
	\label{eq:T0r}
	\begin{aligned}
		\dot{T}_0^{\,r}(t) = f_0^{\,r}(z_0^{\,r}(t), \{T_j^{\,r}(t), q_j^{\,r}(t)\}_{\forall j \in I_0^{\,r}}, T^{\text{ext}}).
	\end{aligned}
\end{equation}
Furthermore, given Assumption \ref{ass:heatingstation}, and since additional bypasses between the supply and the return network are not considered, the heating station water flow, indicated as $q_0$, is equal to the sum of all load flows, i.e., 
\begin{align}
	\label{eq:q0}
	q_0(t) =\sum\limits_{\forall i \in \mathcal{N}_c} q_i^{\,c}(t).
\end{align}
The heating station power is denoted as $P_0$ and it depends on the overall water flow and on the difference between the supply and return temperature, i.e.,
\begin{align}
	\label{eq:P0}
	P_0(t) = c_w q_0(t)(T_0^{\,s}(t)-T_0^{\,r}(t)).
\end{align}

\smallskip
To sum up, it is possible to collect the above described input and output variables into vectors so as to get an overall state-space model of the DHS network. Thus, equations \eqref{eq:Tci}--\eqref{eq:q0} are compacted as 
\begin{equation}
	\label{eq:ss}
	\left\{
	\begin{aligned}
		& \dot{z}(t) = f(z(t), v(t), d(t))\\
		& y(t) = g(z(t), v(t), d(t))  \smash{\text{\;\raisebox{.5\baselineskip}{,}}}
	\end{aligned}
	\right.
\end{equation}
where $z$ is the overall state vector, \mbox{$v = T_0^{\,s}$} is the controllable input, whereas $d =\{P_i^{\,c}\}_{\forall i \in \mathcal{N}_c}$, i.e., the thermal load demands, are the disturbances. Note that, being $T^{\text{ext}}$ and $T_i^{\,c \star}$ constant over time, they are not included as disturbances in the system model. Concerning the outputs, the following ones, being typically measurable, are selected: \mbox{$y = [ T_0^{\,r}, q_0, \{T_i^{\,s}, T_i^{\,c}, q_i^{\,c}\}_{\forall i \in \mathcal{N}_c}']'$}. In detail,  $T_0^{\,r}$ and $q_0$ are needed to compute the heating station thermal power, as evident from \eqref{eq:P0}, whereas the loads supply temperatures, i.e., $T_i^{\,s}$, must be monitored to ensure they respect prescribed operational limits. Finally, as it will be later clarified, the output temperature and water flow of each thermal load, i.e., $T_i^{\,c}$ and $q_i^{\,c}$, are also convenient to be measured.

At the end, considering an appropriate sampling time $\tau_s$, the DHS model \eqref{eq:ss} can be discretized using a suitable integration method. Hence, the discretized system model reads as
\begin{equation}
	\label{eq:ssdisc}
	\left\{
	\begin{aligned}
		& z(k+1) = \tilde{f}(z(k), v(k), d(k)) \\
		& y(k) = \tilde{g}(z(k), v(k), d(k)) \smash{\text{\quad\quad\raisebox{.5\baselineskip}{,}}}
	\end{aligned}
	\right.
\end{equation}
where $k= \lfloor t {/} \tau_s \rfloor$ is the adopted discrete-time index.

\subsection{Case study}
\begin{figure}[t!]
	\centering
	\captionsetup[subfloat]{labelfont=scriptsize,textfont=scriptsize}
	\subfloat[]{ \includegraphics[width=0.5\textwidth]{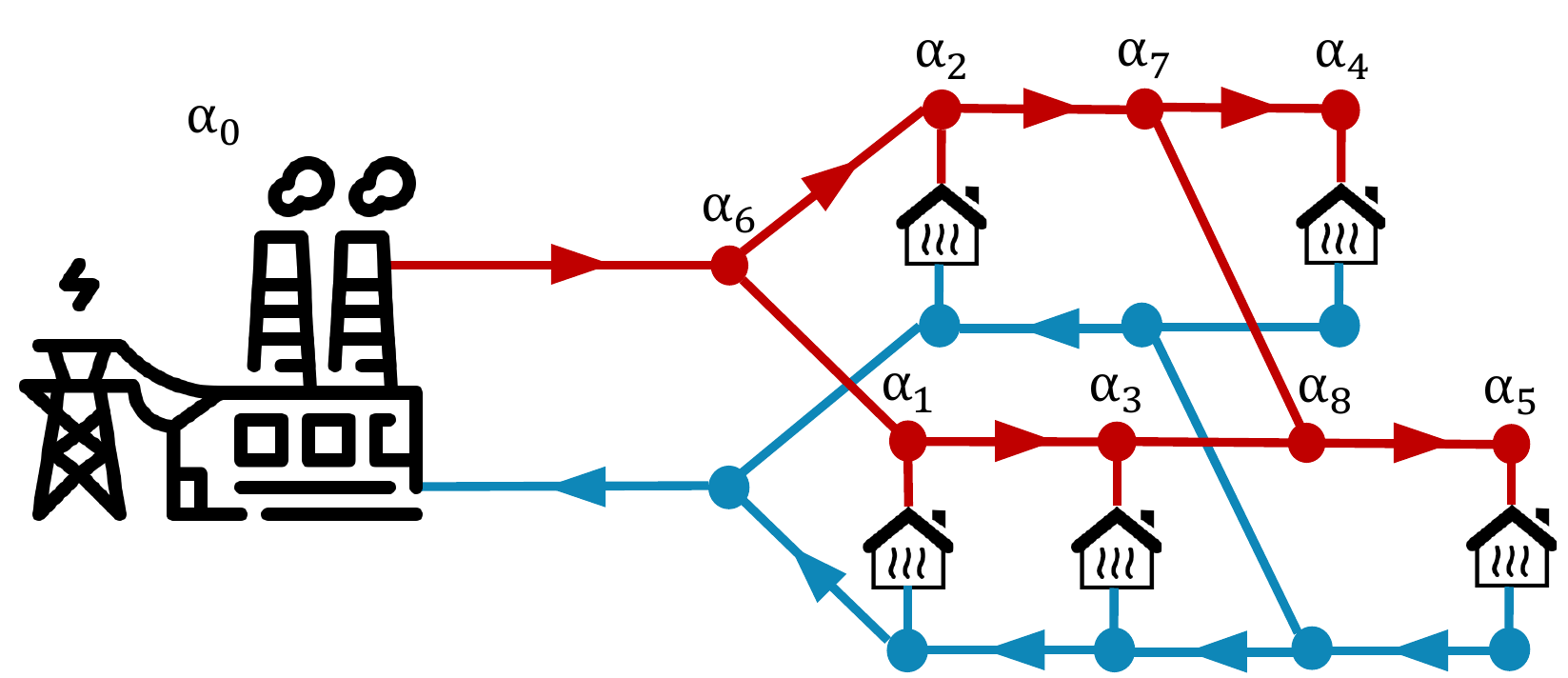} }
	\subfloat[]{\includegraphics[width=0.5\textwidth]{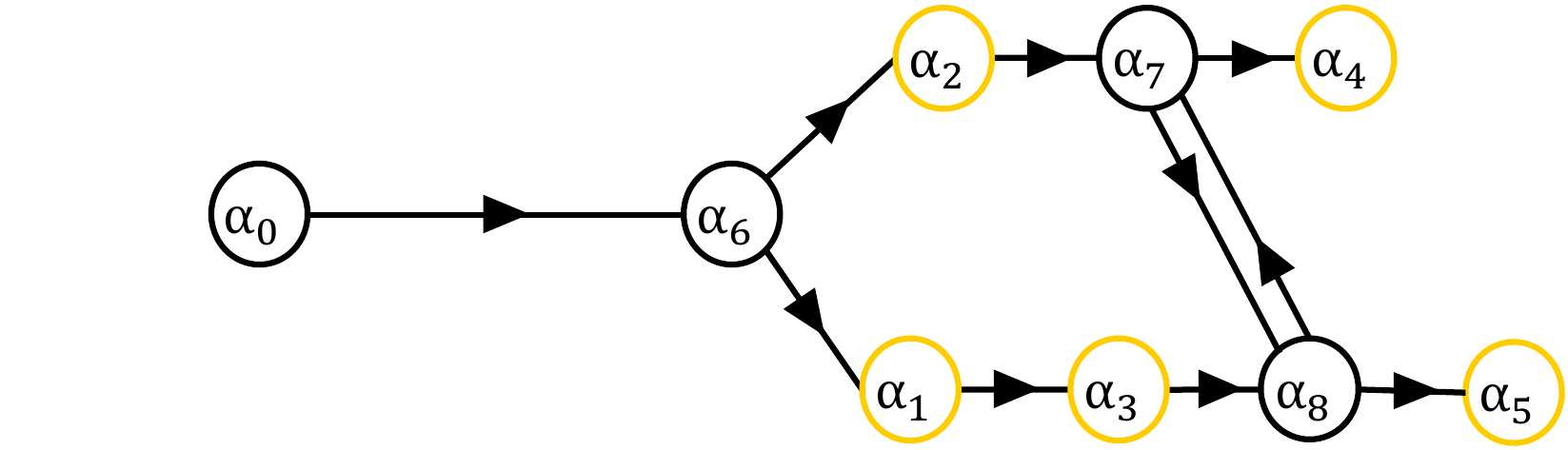} }
	\caption{(a) Schematic representation of the AROMA DHS \cite{krug2021nonlinear}; (b) Graph representation of the AROMA DHS, with load nodes  highlighted in yellow.}
	\label{fig:aroma}
\end{figure}

To better comprehend the proposed modelling method, the considered system benchmark, i.e., the AROMA DHS described in \cite{krug2021nonlinear}, is here briefly introduced. A schematic representation of the AROMA DHS is reported in Figure \ref{fig:aroma}(a). As visible from this scheme, the system is composed of a heating station and a DHS network of nine nodes, including five thermal loads. In particular, the total pipeline length at the supply and return networks is 7262.4 m, whereas other details are available in \cite{krug2021nonlinear}.

As previously discussed, it is possible to define a graph describing the considered DHS, as shown in Figure \ref{fig:aroma}(b). Note that the water flow direction between nodes $\alpha_7$ and $\alpha_8$ may be not determined a priori. Consequently, the corresponding edge is doubled ($e_{78}$ and $e_{87}$), as evident from Figure \ref{fig:aroma}(b). 
For the sake of clarity, as a convention, loads are numbered in increasing order according to their distance along pipelines with respect to the heating station (node $\alpha_0$). Moreover, the nodes which do not represent thermal loads but pure junctions are numbered with indexes greater than the loads ones, again according to their distance from the heating station.

\smallskip
Finally,  the AROMA DHS physical model, described in \cite{krug2021nonlinear}, has been leveraged to develop a dynamic simulator in the Modelica environment \cite{wetter2014modelica}, exploiting the library \cite{DHN4Control}. The developed AROMA DHS simulator will be used in the following both for data collection and for control testing.

\section{Data-based modelling}
\label{sec:id}
As anticipated, for complex large-scale systems such as DHS networks, developing physical models as \eqref{eq:ss} may require a lot of modelling effort and the knowledge of a huge amount of parameters. Therefore, the first objective of the work is to identify  a computationally efficient DHS network model through data-based approaches. The latter deal with the problem of building mathematical models of dynamical systems based on observed data from the plant itself. The procedure to follow is straightforward: input and output signals from the system are collected and processed by a data analysis technique so as to infer a dynamic model \cite{ljung1998system}.

\subsection{Recurrent neural networks}
Various identification techniques can be exploited. Since the system under control is characterized by a nonlinear behaviour, linear models such as AutoRegressive with eXogenous input (ARX) or Output-Error (OE) are not appropriate, as discussed in Section \ref{sec:results}. By contrast, Neural Networks (NNs) \cite{jain1996artificial}, thanks to their enhanced ability to learn nonlinear relationships, are suited to identify complex systems like DHSs. In particular, each NN is characterized by the so-called \textit{hyperparameters}, which include hidden layers and neurons. A hidden layer is an intermediate layer between the NN input and output layer, and it is the collection of neurons which transfer data to layers \cite{karsoliya2012approximating, uzair2020effects}. Within the NNs framework, Recurrent Neural Networks (RNNs) are particularly suited to represent nonlinear dynamical systems and to process time series data \cite{medsker1999recurrent}, being inherently characterized by the presence of state variables \cite{bonassi2022recurrent}. Therefore, RNNs will be exploited in the remaining of the work to identify the DHS network model under investigation. 

\smallskip
In general, RNNs can be described as a dynamical state-space model, i.e.,
\begin{equation}
	\label{eq:rnn}
	\left\{
	\begin{aligned}
		& x(k+1) = \phi(x(k), u(k); \Phi) \\
		& y(k) = \psi(x(k), u(k); \Phi) \smash{\text{\quad\quad\raisebox{.5\baselineskip}{,}}}
	\end{aligned}
	\right.
\end{equation}
where $x \in \mathbb{R}^{n_x}$, $u \in \mathbb{R}^{n_u}$, $y \in \mathbb{R}^{n_y}$ are the state, input and output vectors, respectively. Besides, $\Phi$ is the set of parameters (\textit{weights} and \textit{biases}) of the RNN, which must be tuned during the training procedure \cite{bonassi2022recurrent}. Specifically, a RNN is constituted by $n_l$ hidden layers, each one comprising $n_x^{[i]}$ state variables, with $i=1,\hdots,n_l$, thus implying that the total number of states in \eqref{eq:rnn} is \hbox{$n_x=\sum\limits_{i=1}^{n_l}n_x^{[i]}$}. Note that the number of states of each RNN layer is defined by the selected number of neurons \cite{bonassi2023reconciling}.
With the purpose of identifying the DHS network modelled in \eqref{eq:ssdisc}, the inputs of the RNN model \eqref{eq:rnn} are $u(k) = [v(k)', d(k)'\,]'$.

\smallskip
Despite their potential, RNNs in general do not embed any physical knowledge but they just rely on the available input-output data. Nevertheless, available physical information, such as the network topology in DHSs, is worth to be exploited to enhance their modelling performances.

\subsection{Physics-informed recurrent neural networks} 
\label{sec:pbrnn}
The proposed Physics-Informed Recurrent Neural Networks (PI-RNNs) modelling methodology involves the interconnection of different RNNs according to the physical system structure, so that the so-obtained overall PI-RNN architecture resembles the DHS network topology.

\smallskip
First of all, as later clarified, just a subset of the DHS nodes are of interest from the control perspective, i.e., the one comprising the heating station ($\alpha_0$) and the thermal loads ($\alpha_i$, $\forall i \in \mathcal{N}_c$). Thus, a reduced graph $\tilde{\mathcal{G}}=(\tilde{\mathcal{N}} , \tilde{\mathcal{E}})$ is introduced, where $\tilde{\mathcal{N}}$ denotes the set of these \textit{significant} nodes, i.e., \mbox{$\tilde{\mathcal{N}} = \{0\} \cup \mathcal{N}_c$}. Then, the nodes in $\tilde{\mathcal{N}}$ are interconnected according to their physical dependence with respect to the supply network. In fact, the supply temperature at each load node is influenced by the ones at the inlet load nodes, defined according to the water flow direction. Hence, the set of the edges of this reduced graph, i.e., $\tilde{\mathcal{E}} \subseteq \tilde{\mathcal{N}} \times \tilde{\mathcal{N}}$, is defined as 
	\begin{align}
 		\tilde{\mathcal{E}} =&\big\{(i,j) \, |\, \exists\text{ a path } \{(\beta_1, \beta_2),(\beta_2, \beta_3)\hdots,(\beta_{n-1},\beta_n) \}, \text{with} \nonumber \\ & \beta_1=i, \beta_n=j, (\beta_{k},\beta_{k+1}) \in \mathcal{E}, \, \forall k=\{1,\hdots,n-1\},  \text{and} \nonumber \\ & \beta_k \notin \tilde{\mathcal{N}}, \; \forall k=\{2,\hdots,n-1\} \big\}\,.	\label{eq:reducededge}
	\end{align}
The definition of $\tilde{\mathcal{E}}$ in \eqref{eq:reducededge} expresses the fact that two nodes in $\tilde{\mathcal{N}}$ are connected by an edge if there exists a path in the original DHS graph ${\mathcal{G}}=({\mathcal{N}} , {\mathcal{E}})$  which interconnects them and does not contain any other node in {$\tilde{\mathcal{N}}$}.

\begin{figure}[t]
	\centering
	\captionsetup[subfloat]{labelfont=scriptsize,textfont=scriptsize}
	\subfloat[]{ \includegraphics[width=0.35\textwidth]{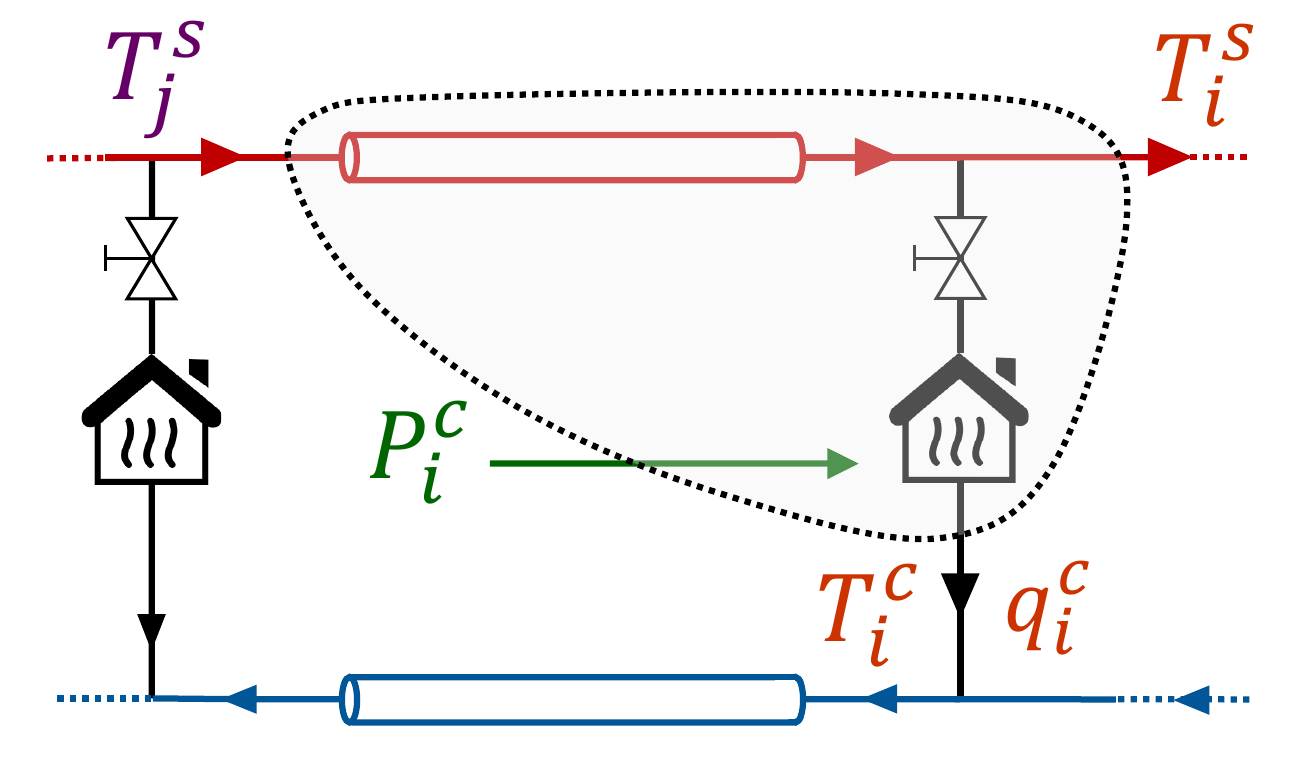} } \hspace{1cm}
	\subfloat[]{\includegraphics[width=0.25\textwidth]{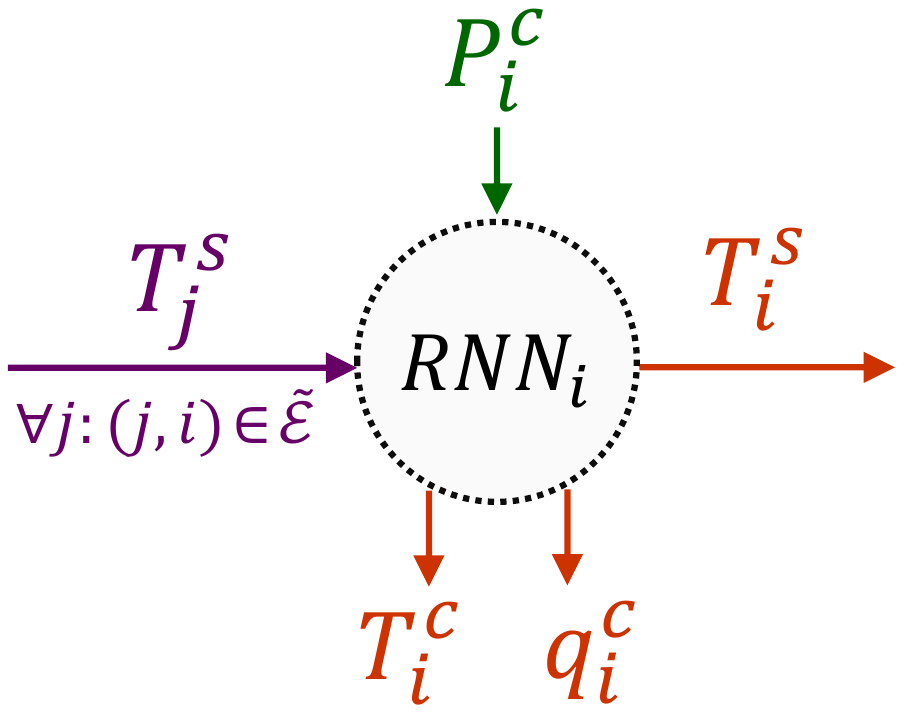} }
	\caption{(a) Schematic representation of the $i$th DHS section comprising the $i$th thermal load and the supply pipe(s) entering node $\alpha_i$ (dotted shadow area); (b)  $i$th load-associated RNN having inputs and outputs paired with the ones of the corresponding $i$th DHS section. }
	\label{fig:loadpbrnn}
\end{figure}

The definition of the reduced graph $\tilde{\mathcal{G}}=(\tilde{\mathcal{N}} , \tilde{\mathcal{E}})$ derives from the fact that, as visible from Figure \ref{fig:loadpbrnn}(a), each load supply temperature is influenced by the ones at the inlet load nodes, defined according to the water flow direction. In particular, let us consider a section of the DHS network comprising the $i$th thermal load node and the supply pipes connecting it with each $j$th inlet node,  $\forall j : (j,i) \in \tilde{\mathcal{E}}$ (dotted shadow area in Figure \ref{fig:loadpbrnn}(a)). It is evident that the supply temperatures of the inlet load nodes, i.e., $\{T_j^{\,s}\}_{\forall j : (j,i) \in \tilde{\mathcal{E}}}$, have a direct impact on the considered $i$th DHS section, and thus they can be modelled as local inputs of this subsystem. The same holds for the thermal demand $P_i^{\,c}$, which acts as an external disturbance significantly influencing the local load water flow and output temperature. On the other hand, the resulting supply temperature $T_i^{\,s}$ can be modelled as an output for the considered $i$th DHS section. This must comply with the load operational limits and it constitutes an input for the subsequent DHS section models, defined based on $\tilde{\mathcal{E}}$. Additionally, the load water flow $q_i^{\,c}$ and the output temperature $T_i^{\,c}$ are modelled as outputs for the $i$th DHS section model as well, since these will be needed to identify the return network dynamics, as explained below.

Consequently, the approach proposed in this work consists in defining a load-associated RNN for each $i$th DHS section, with $i\in \mathcal{N}_c$, comprising the $i$th load and the corresponding supply pipe(s) entering the node,  as depicted in Figure \ref{fig:loadpbrnn}(b). Then, each RNN is interconnected to the others according to the topology of the reduced graph, expressing the dependence among inputs and outputs of the different DHS sections.

\begin{figure}[t]
	\centering
	\captionsetup[subfloat]{labelfont=scriptsize,textfont=scriptsize}
	\subfloat[]{ \includegraphics[width=0.4\textwidth]{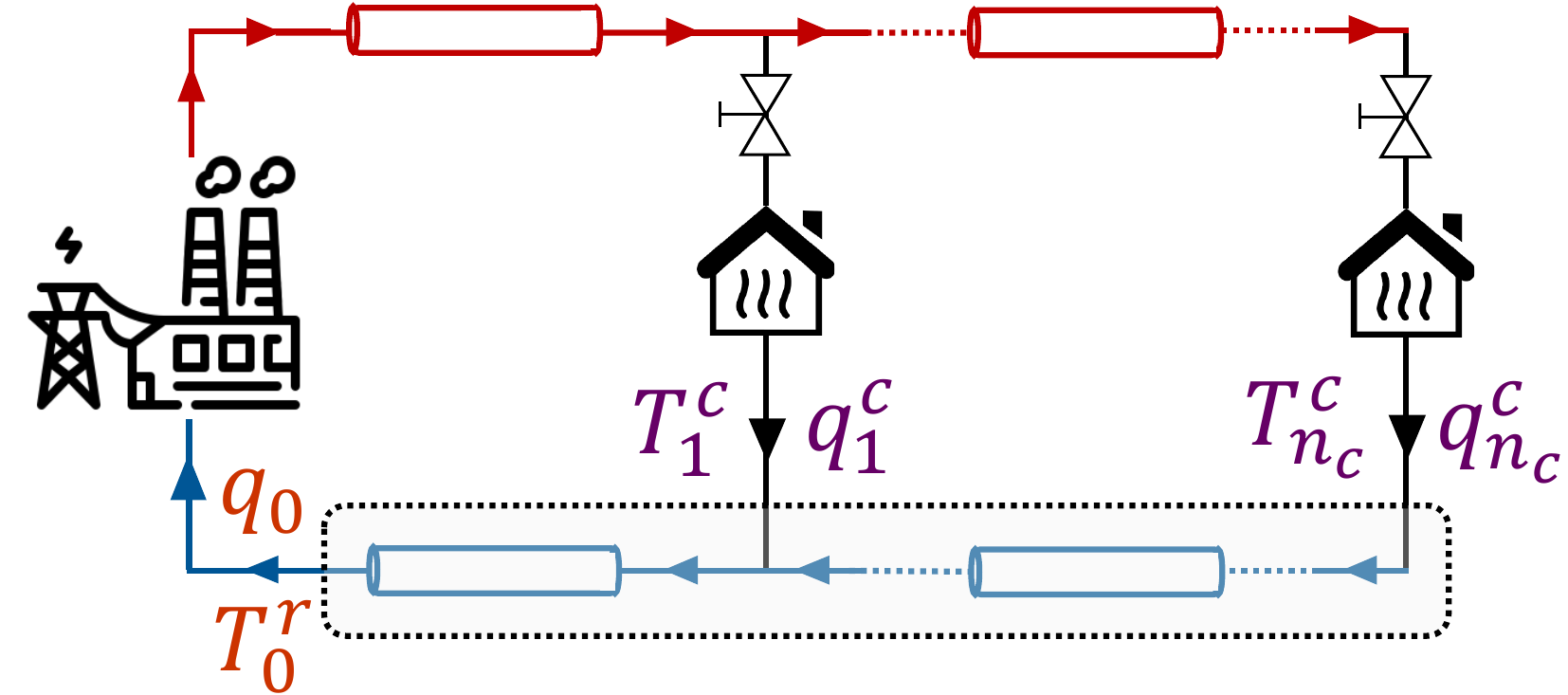} } \hspace{1cm}
	\subfloat[]{\includegraphics[width=0.21\textwidth]{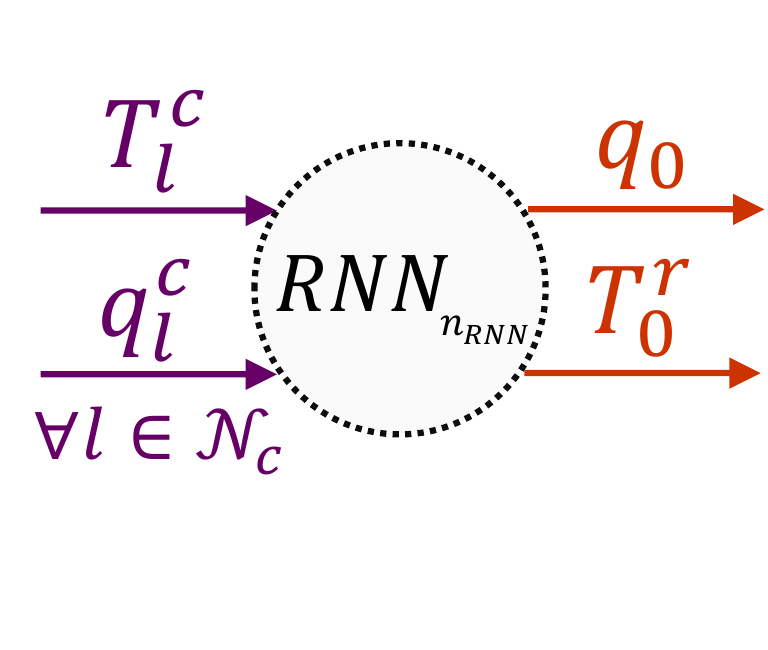} }\hspace{0.8cm}
	\caption{(a) Schematic representation of the return network section (dotted shadow area); (b) Return-associated RNN having inputs and outputs paired with the ones of the corresponding DHS section.}
	\label{fig:returnnetwork}
\end{figure}

On the other hand, a different approach applies for the return network dynamics. Indeed, a single return-associated RNN is employed, since nodal variables at the return network ($T_i^{\,r}$, $q_i^{\,r}$, $\forall i \in \mathcal{N}_c$) are not of interest from the control perspective, as will be evident in Section \ref{sec:MPC}, but only the return temperature $T_0^{\,r}$ and water flow $q_0$ are necessary to compute the heating station produced power $P_0$ in \eqref{eq:P0}. The return-associated RNN receives as inputs the output temperature and water flow of each $i$th load, which are outputs of the $i$th load-associated RNN, \hbox{$\forall \,i \in \mathcal{N}_c$}, and it outputs the return temperature and the water flow at the heating station, i.e., $T_0^{\,r}$ and $q_0$, as evident from Figure \ref{fig:returnnetwork}.

\newpage
Thus, the total number of employed RNNs in the proposed PI-RNN approach is equal to the number of loads plus one, i.e., \hbox{$n_{\text{RNN}}=n_c+1$}, and each $i$th RNN is modelled as
\begin{equation}
	\label{eq:rnni}
	\left\{
	\begin{aligned}
		& x^{[i]}(k+1) = \phi^{[i]}(x^{[i]}(k), u^{[i]}(k); \Phi^{[i]}) \\
		& y^{[i]}(k) = \psi^{[i]}(x^{[i]}(k), u^{[i]}(k); \Phi^{[i]}) \smash{\text{\quad\quad\raisebox{.5\baselineskip}{,}}}
	\end{aligned}
	\right.
\end{equation}
with $i \in \{1,\hdots,n_{\text{RNN}}\}$. 
\\\\\\
The number of states of each $i$th RNN is indicated with $n_x^{[i]}$, i.e., $ x^{[i]} \in \mathbb{R}^{n_x^{[i]}}$, implying that the total number of states of the whole PI-RNN is \hbox{$n_x=\sum\limits_{i=1}^{n_{\text{RNN}}} n_x^{[i]}$}.
In particular, the inputs and outputs of each $i$th RNN are defined as
\\
\\
\begin{itemize}
    \item for each $i$th load-associated RNN, with $i \in\{1,\hdots,n_c\}$:
	\begin{subequations}
		\begin{align}
			& v^{[i]}(k) = \{T_j^{\,s}(k)\}_{\forall j : (j,i) \in \tilde{\mathcal{E}}}, \label{subeq:vi}\\
			& d^{[i]}(k) = P_i^{\,c}(k), \label{subeq:Pci}\\
			& u^{[i]}(k) = [v^{[i]}(k)', d^{[i]}(k)']', \label{subeq:ui} \\[0.25cm]
			& y_s^{[i]}(k) = T_i^{\,s}(k), \label{subeq:ysi}\\
			& y_r^{[i]}(k) = [T_i^{\,c}(k), q_i^{\,c}(k)]', \label{subeq:yri}\\
			& y^{[i]}(k) = [y_s^{[i]}(k)', y_r^{[i]}(k)']', \label{subeq:yi}
		\end{align}
	\end{subequations}
	\\
    \item for the return-associated RNN, with $i = n_{\text{RNN}}=n_c+1$:
	\begin{subequations}
		\begin{align}
			& u^{[i]}(k)=\{T_l^{\,c}(k), q_l^{\,c}(k)\}_{\forall l \in \mathcal{N}_c}, \label{subeq:ur} \\[0.25cm]
			& y^{[i]}(k)=[T_0^{\,r}(k), q_0(k)]'. \label{subeq:yr}
		\end{align}
	\end{subequations}
\end{itemize}

\pagebreak
In other words, each $i$th load-associated RNN, i.e., paired with a DHS section comprising the $i$th load and the supply pipe(s) entering node $\alpha_i$, identifies the corresponding supply temperature among its outputs, see  \eqref{subeq:ysi}, which consequently constitutes an input for the load-associated RNNs influenced by the $i$th one, as evident from \eqref{subeq:vi}, according to the reduced graph interconnections described by $\mathcal{\tilde{E}}$. As shown in Figure \ref{fig:loadpbrnn}(b), each $i$th load-associated RNN returns also as outputs the corresponding load output temperature and water flow, see \eqref{subeq:yri}, which represent an input for the return-associated RNN \eqref{subeq:ur}, as shown in Figure \ref{fig:returnnetwork}(b). Moreover, being the local thermal demand a disturbance, each load-associated RNN is also fed with it,  as evident from \eqref{subeq:Pci}. Finally, the return-associated RNN identifies as outputs the overall water flow and return temperature \eqref{subeq:yr}, being the latter necessary to compute $P_0$ in \eqref{eq:P0}.
Ultimately, by collecting  the variables of all RNNs into vectors, the overall PI-RNN model can be written as in \eqref{eq:rnn} by setting $x = [x^{[1]'},\hdots,x^{[n_{\text{RNN}}]'}]'$, $v = \{v^{[i]}\}_{\forall i : (0,i) \in \tilde{\mathcal{E}}}$, as the supply temperature at the heating station node $\alpha_0$ is the effective external input of the system, $d = [d^{[1]'},\hdots,d^{[n_c]'}]'$, $u = [v', d']'$, and $y = [y^{[1]'},\hdots,y^{[n_{\text{RNN}}]'}]'$. \\[0.1cm]

\subsubsection*{PI-RNN modelling applied to the AROMA DHS}
\phantom{a}\\[-0.3cm]

\begin{figure}[b]
	\centering
	\captionsetup[subfloat]{labelfont=scriptsize,textfont=scriptsize}
	\subfloat[]{ \includegraphics[width=0.5\textwidth]{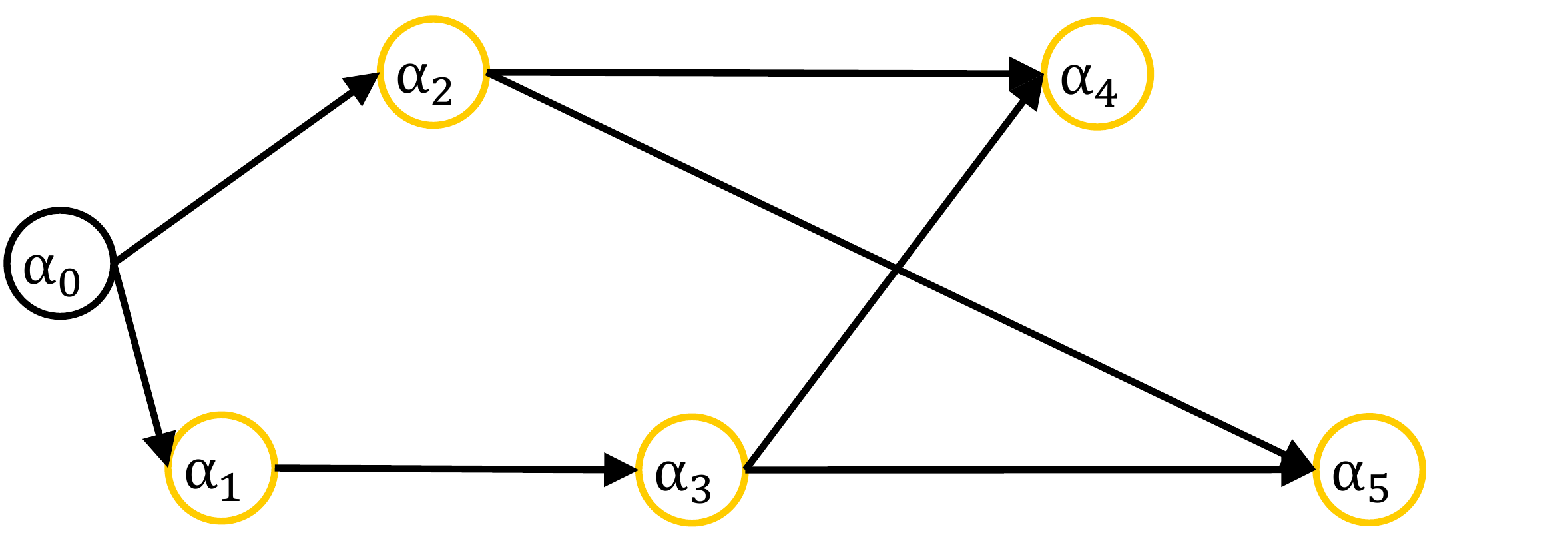} } 
	\subfloat[]{\includegraphics[width=0.5\textwidth]{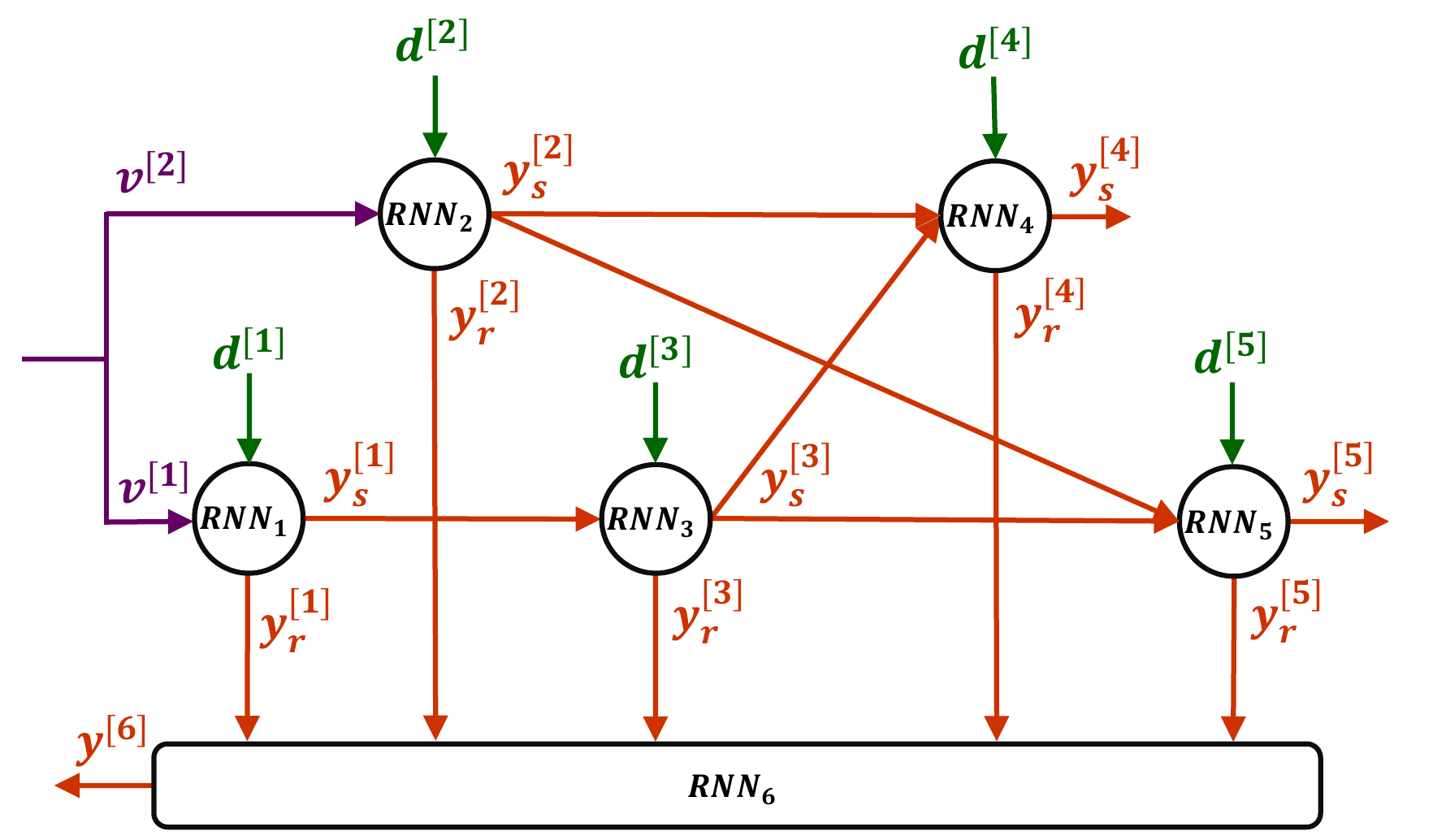} }
	\caption{(a) AROMA DHS reduced graph: nodes containing loads are highlighted in yellow; (b) AROMA DHS PI-RNN architecture: inputs are depicted in purple, disturbances in green and outputs in orange.}
	\label{fig:pbrnntot}
\end{figure}

As discussed, starting from the physical topology of the AROMA DHS, reported in Figure \ref{fig:aroma}(a), one can extract the oriented graph depicted in Figure \ref{fig:aroma}(b). Therefore, following the procedure described in the previous section, the reduced graph shown in Figure \ref{fig:pbrnntot}(a) can be defined, which represents how load supply temperatures influence each other. Finally, the proposed \mbox{PI-RNN} architecture, which reflects the physical system topology, is encoded according to the information contained in the reduced graph, as shown in Figure \ref{fig:pbrnntot}(b). Please note that $v=v^{[1]} = v^{[2]} = T_0^{\,s}$,  being the supply temperatures at nodes $\alpha_1$ and $\alpha_2$ directly affected by the one at the heating station node $\alpha_0$, which is the overall system input. Moreover, $v^{[3]} = y_s^{[1]}$, being $\alpha_1$ the only preceding significant node for $\alpha_3$, $v^{[4]} = v^{[5]} = [y_s^{[2]'},y_s^{[3]'}]'$, being $\alpha_2$ and $\alpha_3$ the preceding significant nodes for $\alpha_4$ and $\alpha_5$, whereas $u^{[6]}=[y_r^{[1]'},y_r^{[2]'},y_r^{[3]'},y_r^{[4]'},y_r^{[5]'}]'$, since the return-associated RNN is fed with the load output temperatures and water flows which are outputs for the five load-associated RNNs. 
\smallskip
\begin{remark}
	\label{remark:loadcluster}
	Considering the AROMA DHS case study, a different RNN is paired with each load. However, in case DHSs are characterized by numerous thermal loads, these can be grouped as shown in \cite{chicco2012overview}. In this way, a single RNN can be used to model each loads cluster, thus limiting the size of the overall PI-RNN model.
\end{remark}
\smallskip
\begin{remark}
	\label{remark:loadorder}
	Thanks to the matching between the PI-RNN architecture and the DHS network topology, it is possible to discriminate which RNNs are characterized by identification issues in modelling the corresponding physical variables and, hence, which should be aided through a large number of neurons to boost their modelling performance. For instance, the identification error may increase for the RNNs paired with the farthest DHS sections from the heating station, due to growing uncertainty related to heat and head losses over water pipelines. Thus, the larger the distance between the DHS section and the heating station, the higher the number of neurons, and consequently the number of states, of the corresponding RNN.
\end{remark}
\smallskip
\begin{remark}
	\label{remark:pci}
    In the PI-RNN approach, each RNN can be fed with some additional physical knowledge to improve its modelling performance. For instance, in order to give the information regarding the total DHS thermal demand to each $i$th load-associated RNN, the sum of the other thermal demands can be also provided, i.e., $d^{[i]}(k)=[P_i^{\,c}(k),\,P^{\,c}_{\text{tot},i}(k)]'$, with \mbox{$P^{\,c}_{\text{tot},i} (k)=\!\! \sum\limits_{\forall l \in \mathcal{N}_c, l \neq i} P_l^{\,c} (k)$}.
\end{remark}
\pagebreak
\section{Nonlinear Model Predictive Control}
\label{sec:MPC}
Before showing the performances achieved by the proposed PI-RNN modelling approach, an NMPC regulation strategy is formulated, which periodically optimizes the DHS operation exploiting the derived RNN-based dynamical models. 

\smallskip
Let us consider a sampling period $\tau_s$ and a prediction horizon of $N$ steps. Thus, leveraging the \textit{receding horizon} strategy \cite{maciejowski2002predictive}, the following NMPC problem is solved at each time instant $t=k_s\tau_s$, with $k_s \in \mathbb{N}$,
\begin{subequations}
	\label{eq:mpc}
	\begin{align}
		& \min_{T_0^{\,s}({\cdot})} \sum_{k = k_s}^{k_s+N-1} (c_{el}(k) { P_{0}(k)}/{\eta}) +  c_t \sum_{i=1}^{n_c} (T_i^{\,s}(N)-T^{\star})^2 \label{subeq:costf} \\
		& \textrm{subject to}, \forall \; k \in \{k_s,\hdots,k_s+N-1\}, \nonumber \\[0.2cm]
		& \qquad x(k+1) = \phi(x(k), u(k); \Phi), \label{eq:subeqxkp1}\\
		& \qquad y(k) = \psi(x(k), u(k); \Phi), \label{eq:subeqy} \\[0.2cm]
		& \qquad x(k_s) = \hat{x}_0, \label{subeq:obs} \\[0.2cm]
		& \qquad \underline{T}_0^{\,s} \leq T_0^{\,s}(k) \leq \overline{T}_0^{\,s}, \label{eq:subeqT0s}\\
		& \qquad \underline{T}_0^{\,r} \leq T_0^{\,r}(k) \leq \overline{T}_0^{\,r}, \label{eq:subeqT0r}\\[0.2cm]
		& \qquad P_0(k) = c_w q_0(k)(T_0^{\,s}(k)-T_0^{\,r}(k)), \label{subeq:P0eq}
		\\& \qquad \underline{P}_0 \leq P_0(k) \leq \overline{P}_0, \label{eq:subeqP0} \\[0.2cm]
		& \qquad \underline{T}_i^{\,s}(k) \leq T_i^{\,s}(k) \leq \overline{T}_i^{\,s}(k), \; \forall i \in \mathcal{N}_c,  \label{eq:subeqTls} \\[0.2cm]
		& \qquad -\overline{\Delta T}_{0}^{\,s} \leq  T_0^{\,s}(k+1)- T_0^{\,s}(k)\leq \overline{\Delta T}_{0}^{\,s}, \label{eq:subeqDeltaT0s}\\
		& \qquad T_0^{\,s}(k) = {T}_0^{\,s}(\lfloor k/N_b \rfloor \cdot N_b). \label{eq:blocking}	
	\end{align}
\end{subequations}

\smallskip
In detail, the cost function \eqref{subeq:costf} minimizes the production cost of the heating station: the power $P_0$ is multiplied by the time-varying price $c_{el}$ and divided by the heating station thermal efficiency $\eta$. Moreover, a terminal cost is added in the cost function, weighted via $c_t$, to discourage significant variations of the load supply temperatures from a nominal reference value $T^{\star}$. 

\smallskip
The dynamical model of the DHS network is embedded in the NMPC formulation in \eqref{eq:subeqxkp1}-\eqref{eq:subeqy}, reported in the generic form of \eqref{eq:rnn}, so as to include either the standard RNN or the proposed PI-RNN model. It is worth noting that, in principle, \eqref{eq:subeqxkp1}-\eqref{eq:subeqy} could be replaced by the DHS network physical model, i.e., \eqref{eq:ssdisc}, which, as discussed, leads to a large-scale optimization problem hard to be solved \cite{krug2021nonlinear}. 
Independently of the selected model, as evident from \eqref{subeq:obs}, the system state $x$ must be initialized at each NMPC iteration with $\hat{x}_0$, supposed to be measured or estimated. In fact, being the state typically not accessible in RNN models, state observers could be necessary, e.g., the ones proposed in
\cite{bonassi2021nonlinear, bonassi2022recurrent}.
Constraints \eqref{eq:subeqT0s} and \eqref{eq:subeqT0r} are included to comply with temperature limits at the heating station, whose produced thermal power is modelled in \eqref{subeq:P0eq}  and bounded in \eqref{eq:subeqP0}. 
Moreover, constraint \eqref{eq:subeqTls} is imposed to guarantee that the supply temperature at each thermal load respects prescribed limits, enabling the proper heat delivery and functioning of local load exchangers.
In particular, note that these temperature limits may change over time, e.g., being higher by day and lower by night, consistently with the thermal demand daily trend \cite{la2023optimal}. Moreover, being \eqref{eq:subeqT0r}, \eqref{eq:subeqP0}, and \eqref{eq:subeqTls} imposed on output variables, in order to ensure problem feasibility \cite{maciejowski2002predictive}, these should be stated as \textit{soft} constraints by means of slack variables, which, for the sake of clarity, are not here explicitly reported. \\
Finally, to reduce the computational complexity induced by the nonlinear model \eqref{eq:subeqxkp1}-\eqref{eq:subeqy}, constraints \eqref{eq:subeqDeltaT0s} and \eqref{eq:blocking} are added. The former implies that the variation of the heating station supply temperature between two consecutive time instants is limited by $\overline{\Delta T}_{0}^{\,s}>0$. Constraint \eqref{eq:blocking} is commonly referred to as input blocking strategy \cite{1430345}, since it limits control variables to vary every $N_b\ll N$ steps over the prediction horizon, reducing the problem degrees of freedom. However, being the NMPC regulator executed with a period $\tau_s$, the manipulated input, i.e., $T_0^s$, will still vary at each $t=k_s \tau_s$. Please note that constraints \eqref{eq:subeqDeltaT0s} and \eqref{eq:blocking} are not necessary from a conceptual point of view, but they enable to adopt larger prediction horizons, which can be necessary to effectively optimize DHSs, given their slow dynamical transients.

\smallskip
Overall, the just described optimization problem constitutes a basic example, since more advanced NMPC strategies for DHSs are out of the scope of this article. For instance, a thermal energy storage (TES) could be considered in the heating station modelling as a further degree of freedom.

\section{Numerical results}
\label{sec:results}
In this section, the performances of the proposed modelling and control approaches applied to the AROMA DHS benchmark are presented.

\subsection{Identification results}
First, to properly identify the system dynamics, a significant dataset of input-output samples is collected. Therefore, the system inputs, i.e., the supply temperature $T_0^{\,s}$ and the thermal demands $P_i^{\,c}$, $\forall i \in \mathcal{N}_c$, are varied using multilevel pseudorandom binary sequences (MPRBS), composed of steps of different amplitudes and interval time sizes. Considering the system transients, which are typically slow in DHSs, it is reasonable to collect data with a sampling time \hbox{$\tau_s = 5$ min}. The system is thus simulated to gather a dataset $\mathcal{D}_{\text{tot}}$ of $15690$ samples (properly split in training, validation, and testing sets, denoted as $\mathcal{D}_{\text{train}}$, $\mathcal{D}_{\text{val}}$, $\mathcal{D}_{\text{test}}$, respectively). Let us recall that the outputs of interest of the AROMA DHS are the five loads supply and output temperatures and their absorbed water flows (respectively $T_i^{\,s}$, $T_i^{\,c}$ and $q_i^{\,c}$, $\forall i \in \mathcal{N}_c$), as well as the overall return temperature $T_0^{\,r}$ and water flow $q_0$.

\smallskip
Second, in order to quantitatively evaluate the identification performances, specific performance indexes are defined. In particular, the FIT index and the coefficient of determination $R^2$ are employed, reported in equations \eqref{eq:fit} and \eqref{eq:r2}, respectively. The FIT index assesses the model overall accuracy on the test set, and it defined as
\begin{equation}
	\label{eq:fit}
	\begin{aligned}
		& \text{FIT} = \left( 1- \frac{\lVert \vec{y}_{\text{test}}-\vec{\hat{y}}_{\text{test}} \rVert_2}{\lVert\ \vec{y}_{\text{test}}-\mathbb{1}' \otimes y_{\text{test}}^{\text{avg}}\rVert_2} \right) \cdot 100,
	\end{aligned}
\end{equation}
where \hbox{$\vec{\hat{y}}_{\text{test}} = \big [\{\hat{y}'(i)\}_{\forall i \in \mathcal{D}_{\text{test}}} \big ]'$} is the sequence of identified outputs, \hbox{$\vec{y}_{\text{test}} = \big [\{y'(i)\}_{\forall i \in \mathcal{D}_{\text{test}}}\big ]'$} is the sequence of measured ones and $y_{\text{test}}^{\text{avg}}$ is its average, i.e., defined as \hbox{$y_{\text{test}}^{\text{avg}} = \frac{1}{|\mathcal{D}_{\text{test}}|} \sum\limits_{\forall i \in \mathcal{D}_{\text{test}}} \vec{y}_{\text{test}}(i)$}, as discussed in \cite{bonassi2021nonlinear}.\\
The $R^2$ index is leveraged to assess the modelling accuracy of each identified output with respect to the test set \cite{barrett1974coefficient}. The $R_j^2$ related to the $j$th output is defined as 
\begin{equation}
	\label{eq:r2}
	\begin{aligned}
		& R_j^2 = \left( 1- \frac{\sum\limits_{\forall i \in \mathcal{D}_{\text{test}}} (y_{j}(i)-\hat{y}_{j}(i))^2}{\sum\limits_{\forall i \in \mathcal{D}_{\text{test}}} \!\!(y_{j}(i)-y_{j}^{\text{avg}})^2} \right) \cdot 100,
	\end{aligned}
\end{equation}
where \hbox{$y_{j}^{\text{avg}} = \frac{1}{|\mathcal{D}_{\text{test}}|} \sum\limits_{\forall i \in \mathcal{D}_{\text{test}}} {y}_{j}(i)$} for each $j$th output. In particular, the minimum coefficient of determination, i.e., \mbox{$\underline{R}^2 = \min\limits_{j=1,\hdots,n_y} R_j^2$}, related to the output identified with the worst accuracy and the maximum one, i.e., \mbox{$\overline{R}^2=\max\limits_{j=1,\hdots,n_y} R_j^2$}, related to the output identified with the best accuracy are evaluated to assess the modelling performances of the developed data-based models.

\smallskip
As anticipated, linear models such as State-Space (SS), AutoRegressive with eXogenous input (ARX) and Output-Error (OE) are not able to properly model the system dynamics, yielding very low FIT values, and thus they are not considered further. Consequently, standard RNNs are tested in the first place, and then PI-RNNs are developed and compared to the former. The implementation of both types of NNs is performed with the Python programming language (version 3.10), using the library developed in \cite{ssnet} for standard RNNs and customizing it to build up PI-RNNs. The training procedure employed for the different RNNs exploits the so-called Truncated Back-Propagation Through Time (TBPTT) method, thoroughly described in \cite{bonassi2023reconciling}. Ultimately, all computations are carried out on a laptop with an Intel Core i7-11850H processor.

\smallskip
In detail, two families of RNN architectures are first tested, i.e., Long Short-Term Memory (LSTM) \cite{10178405} and Gated Recurrent Unit (GRU) \cite{bonassi2021nonlinear}, as well as different combinations of hyperparameters, i.e., amount of hidden layers, of neurons and optimizers (e.g., ADAM and RMSProp \cite{kingma2014adam}). By comparing the performance indexes of the different combinations of NNs and hyperparameters (FIT, $\underline{R}^2$, $\overline{R}^2$, best epoch and training time to reach the best epoch), it turns out that GRU NNs are the most suitable to identify the AROMA DHS model.

\smallskip
Thus, the performances of a standard GRU and of a Physics-Informed GRU model (PI-GRU) are now compared. These are  trained with the 15690-sample dataset over 1500 epochs, with the ADAM optimizer and a learning rate of 0.003, so as to get a good trade-off between convergence speed and excessive oscillations avoidance \cite{bemporad2022recurrent}. 

Since the AROMA DHS is composed of five loads, the PI-GRU model is composed of six GRUs ($n_{\text{RNN}}=n_c+1$, see Figure \ref{fig:pbrnntot}(b)), each one implemented with a single hidden layer. To make a fair comparison, the standard GRU is composed of six hidden layers accordingly ($n_l = n_{\text{RNN}}$).\\
Moreover, as highlighted in Remark \ref{remark:loadorder}, the number of neurons of the load-associated GRUs in the PI-GRU model increases with the distance of the corresponding thermal load from the heating station. On the other hand, the return-associated GRU, being paired with a DHS section comprising the overall return network, is assigned a large number of neurons as well. Therefore, given that in GRU networks the state dimension matches the number of neurons of each layer \cite{bonassi2023reconciling}, $n_x^{[i]}$ increases with $i$, as load nodes are numbered according to their distance from the heating station, as previously discussed. By contrast, since hidden layers do not have a physical matching in the standard GRU model, the same amount of neurons is set for each hidden layer, so that its total number of states coincides with the PI-GRU one.
\\
Then, as highlighted in Remark \ref{remark:pci}, each load-associated PI-GRU is fed, among other inputs described in Section \ref{sec:pbrnn}, with the cumulative power consumption.

\begin{figure}[t!]
	\centering
\includegraphics[width=0.6\textwidth]{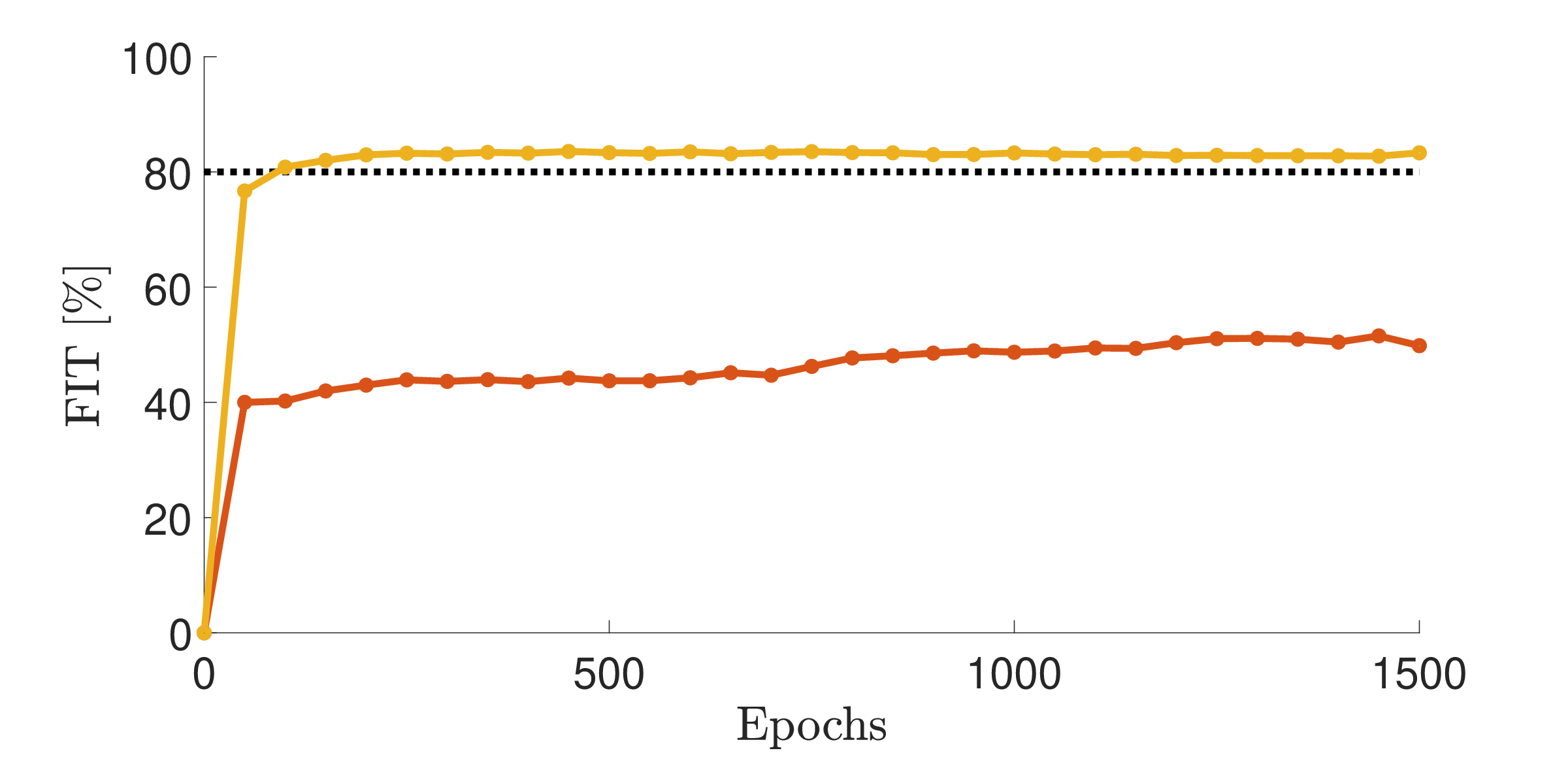}
	\caption{Comparison between the FIT trend of a standard GRU (orange) and of a PI-GRU (yellow) over the training procedure, both having 54 states and trained with a 15690-sample dataset. The 80\% FIT is depicted in dotted black.}
	\label{fig:fit54neuroni}
\end{figure}

Figure \ref{fig:fit54neuroni} reports the comparison between the FIT trend of a 54-state standard GRU and of a PI-GRU one, for the AROMA DHS benchmark. Specifically, $n_x = 54$ is chosen as it is the minimum amount of states that enables the standard GRU to reach approximately a FIT of 50\%. The PI-GRU takes 371 epochs and a training time of 61 minutes to reach its best FIT value of 83.3\%, whereas the standard GRU takes 1171 epochs and a training time of 106 minutes to reach its best FIT value of 49.8\%, as evident from Figure  \ref{fig:fit54neuroni}. For the sake of completeness, the identification procedure is repeated multiple times due to the random initialization of RNNs weights and biases and the random subsequences extraction of the TBPTT method. The obtained results are reported in Table \ref{table:fit54neuroni}, where the average FIT, $\underline{R}^2$ and $\overline{R}^2$ values are shown, together with their standard deviation. 
These results show how the PI-GRU outperforms the standard GRU, even though the two networks are characterized by the same hyperparameters. In particular, the enhancement in the FIT and in  $\underline{R}^2$ reported in Table \ref{table:fit54neuroni} is promising. Moreover, the PI-GRU takes less than 100 epochs to exceed a FIT of 80\%, which is a value that the standard GRU does not even reach within 1500 epochs (see Table \ref{table:fit54neuroni}). 

\begin{table}[t!]
	\centering
	\caption{Comparison between the performance of a standard GRU and a PI-GRU, both having 54 states and trained with 15690 samples.}
	\begin{tabular}{c|ccccc}
		& $n_x$ &$n_x^{[i]}$ & FIT [\%] & $\underline{R}^2$ [\%]& $\overline{R}^2$ [\%]\\ [0.1cm] \hline  \\[-6px] 
		GRU & 54&[9,9,9,9,9,9] & 51.0$\pm$2.5 & 30.3$\pm$5.2 & 96.5$\pm$0.9 \\[2px] 
		PI-GRU & 54&[6,6,6,8,12,16] & 83.4$\pm$0.1 & 87.9$\pm$0.3 & 99.3$\pm$0.1
	\end{tabular}
	\label{table:fit54neuroni}
\end{table}

\begin{figure}[t!]
	\centering	\captionsetup[subfloat]{labelfont=scriptsize,textfont=scriptsize}
	\subfloat[]{ \includegraphics[width=0.4\textwidth]{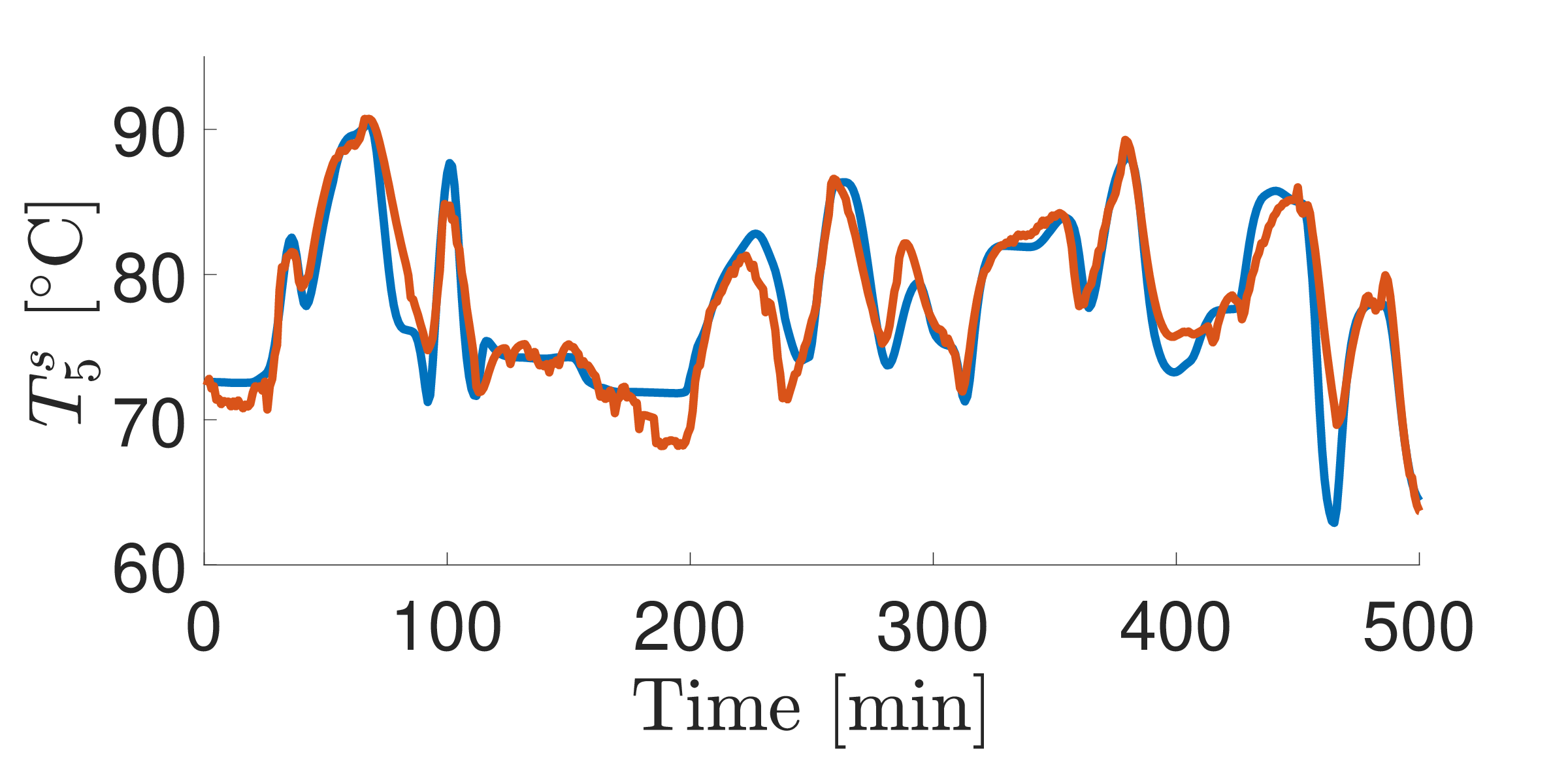} }\hspace{0.2cm}
	\subfloat[]{\includegraphics[width=0.4\textwidth]{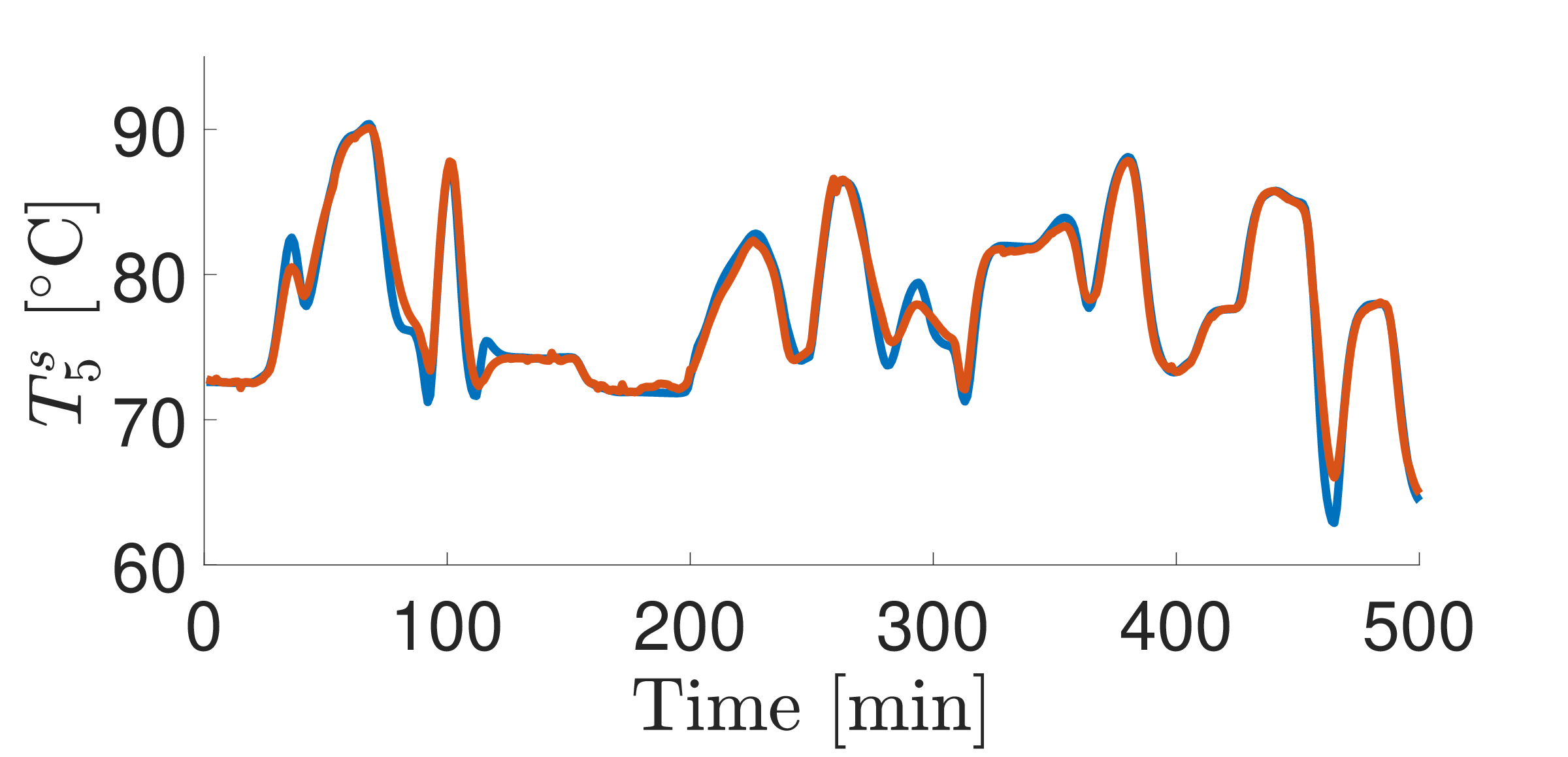} } \\
	\subfloat[]{ \includegraphics[width=0.4\textwidth]{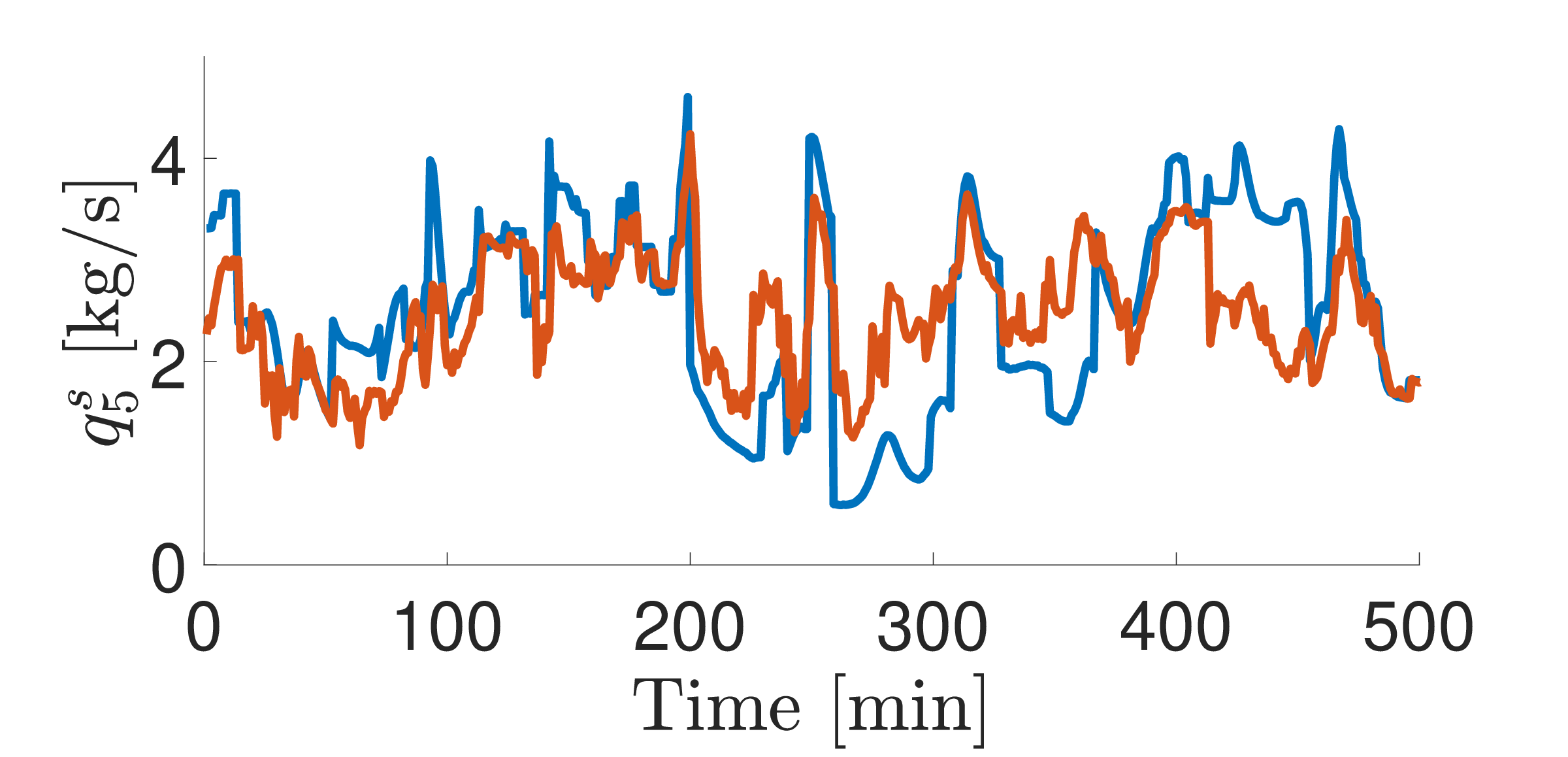} }\hspace{0.2cm}
	\subfloat[]{\includegraphics[width=0.4\textwidth]{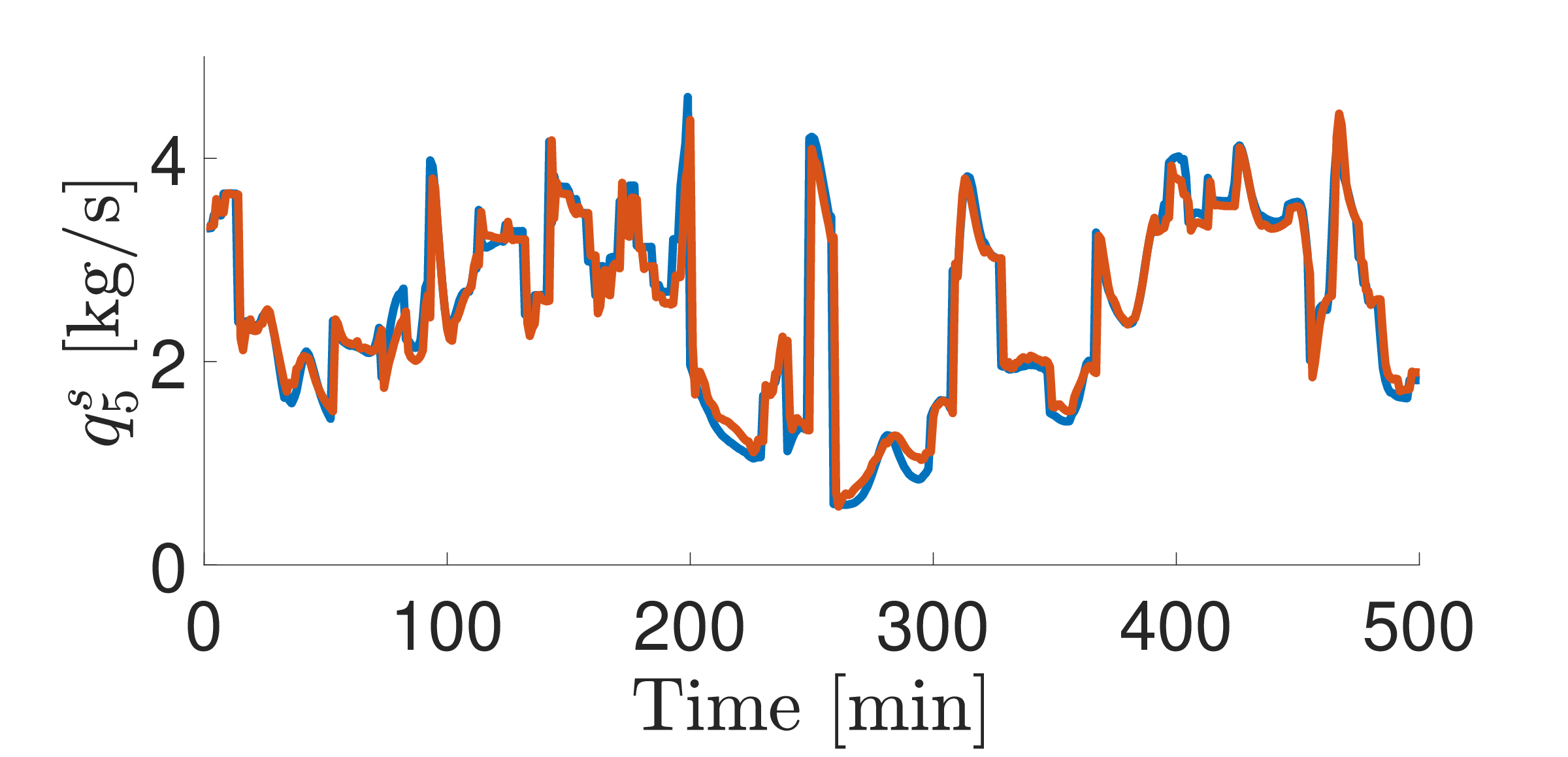} }
	\caption{GRU and PI-GRU identification results. The identified variable is depicted in orange, the measured one in blue. (a) $T_5^{\,s}$ identified by the standard GRU; (b) $T_5^{\,s}$ identified by the PI-GRU; (c) $q_5^{\,c}$ identified by the standard GRU; (d) $q_5^{\,c}$ identified by the PI-GRU.}
	\label{fig:trend}
\end{figure}

Moreover, in Figure \ref{fig:trend}, the measured supply temperature and water flow trend of the load placed in node $\alpha_5$ is compared with the predictions both of the 54-state standard GRU and of the PI-GRU one. 
Being $\alpha_5$ the farthest node from the heating station, only the PI-GRU is able to properly identify both $T_5^{\,s}$ and $q_5^{\,c}$, whereas the standard GRU commits a considerable modelling error (see Table \ref{table:fit54neuroni}).
In fact, the variables paired with the farthest loads are the most challenging to be identified, achieving low $R^2$ values. {This identification local issue, however, can be tackled by PI-GRUs through a suitable choice of the number of neurons, as discussed in Remark \ref{remark:loadorder}, but it cannot be tackled by standard GRUs, and RNNs in general, as their architecture does not have a physical interpretation.}

\begin{figure}[t]
	\centering
	\captionsetup[subfloat]{labelfont=scriptsize,textfont=scriptsize}
	\subfloat[]{ \includegraphics[width=0.4\textwidth]{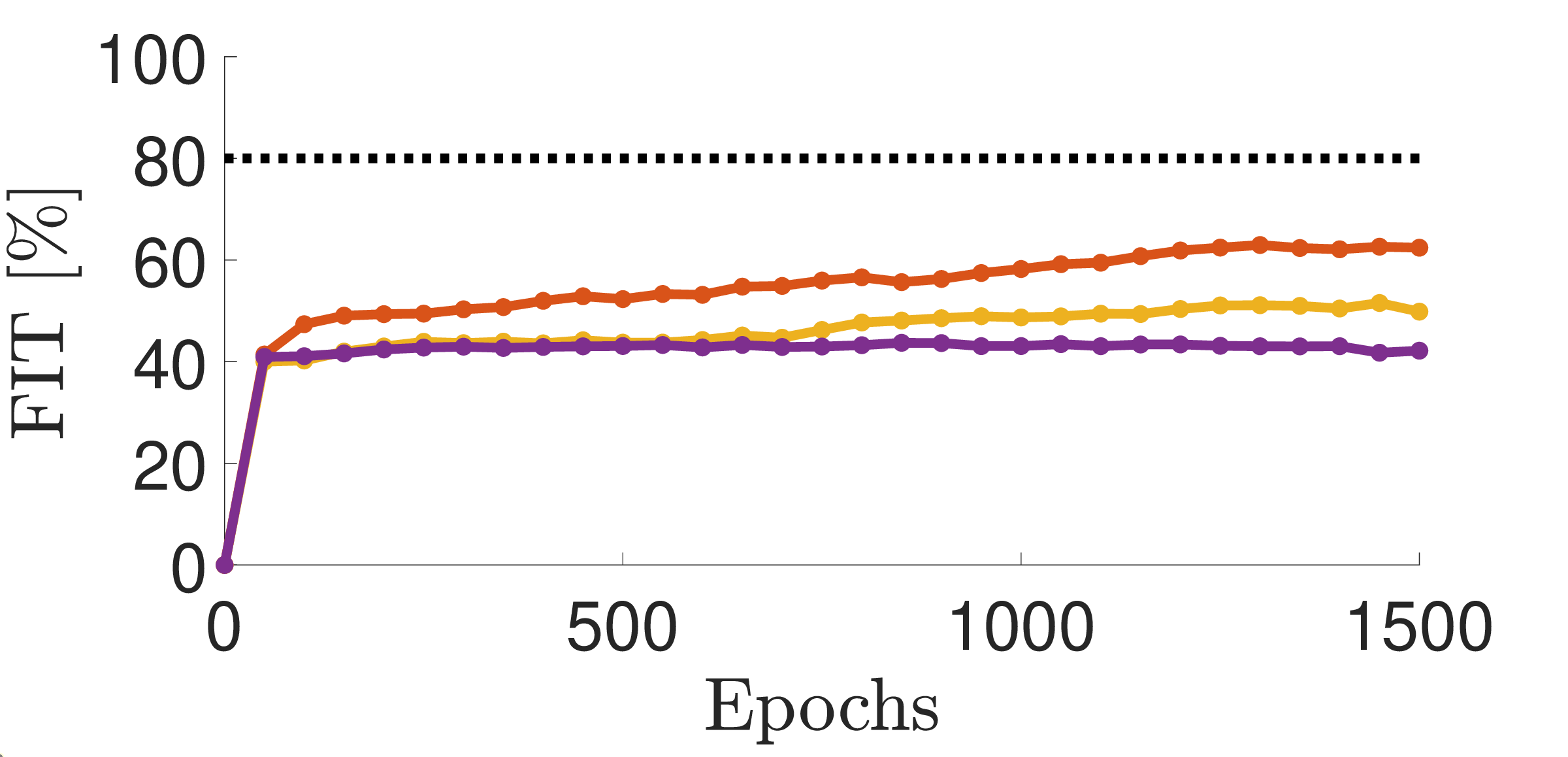} }\hspace{0.3cm}
	\subfloat[]{\includegraphics[width=0.4\textwidth]{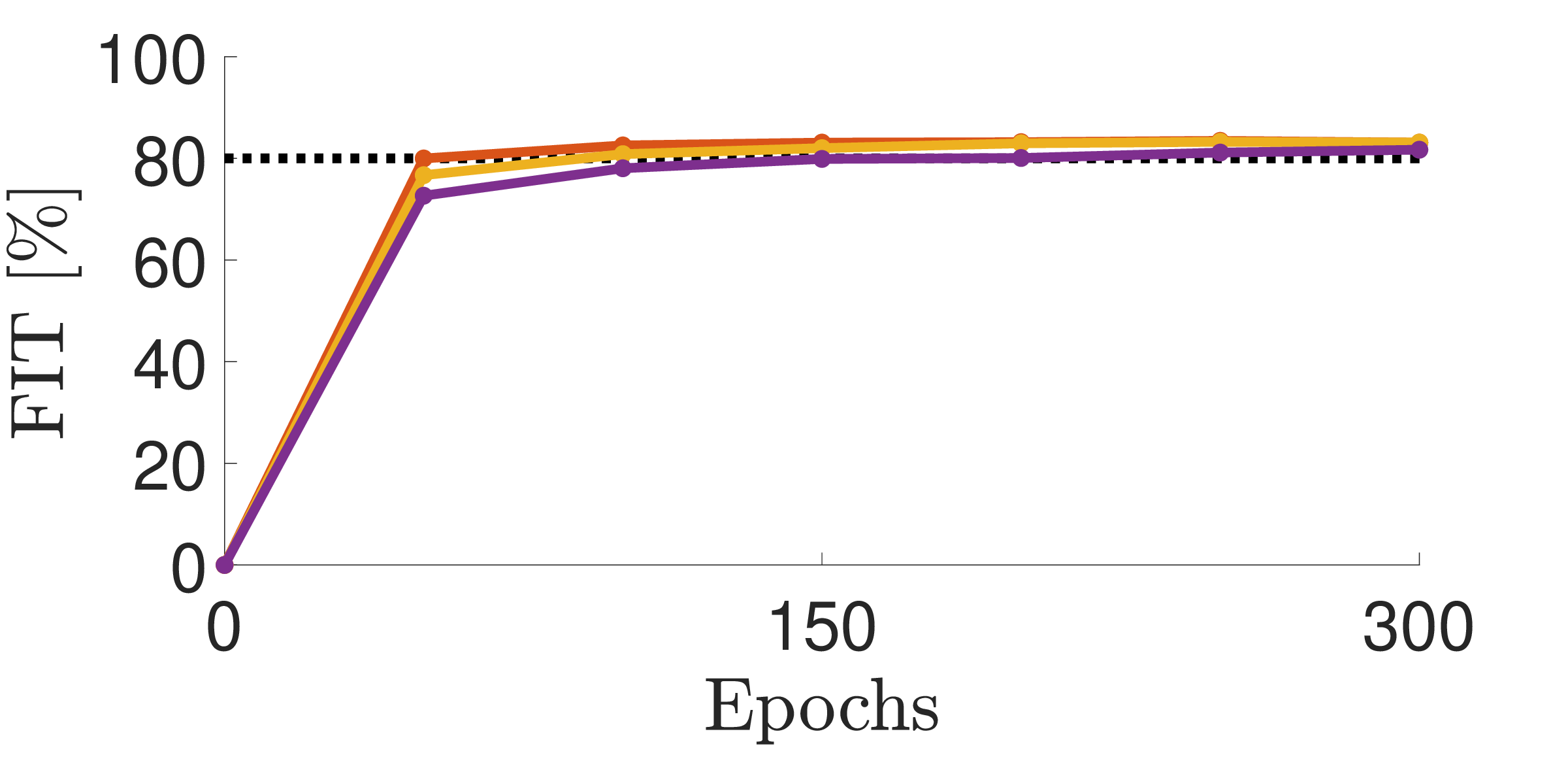} } \\
	\subfloat[]{\includegraphics[width=0.4\textwidth]{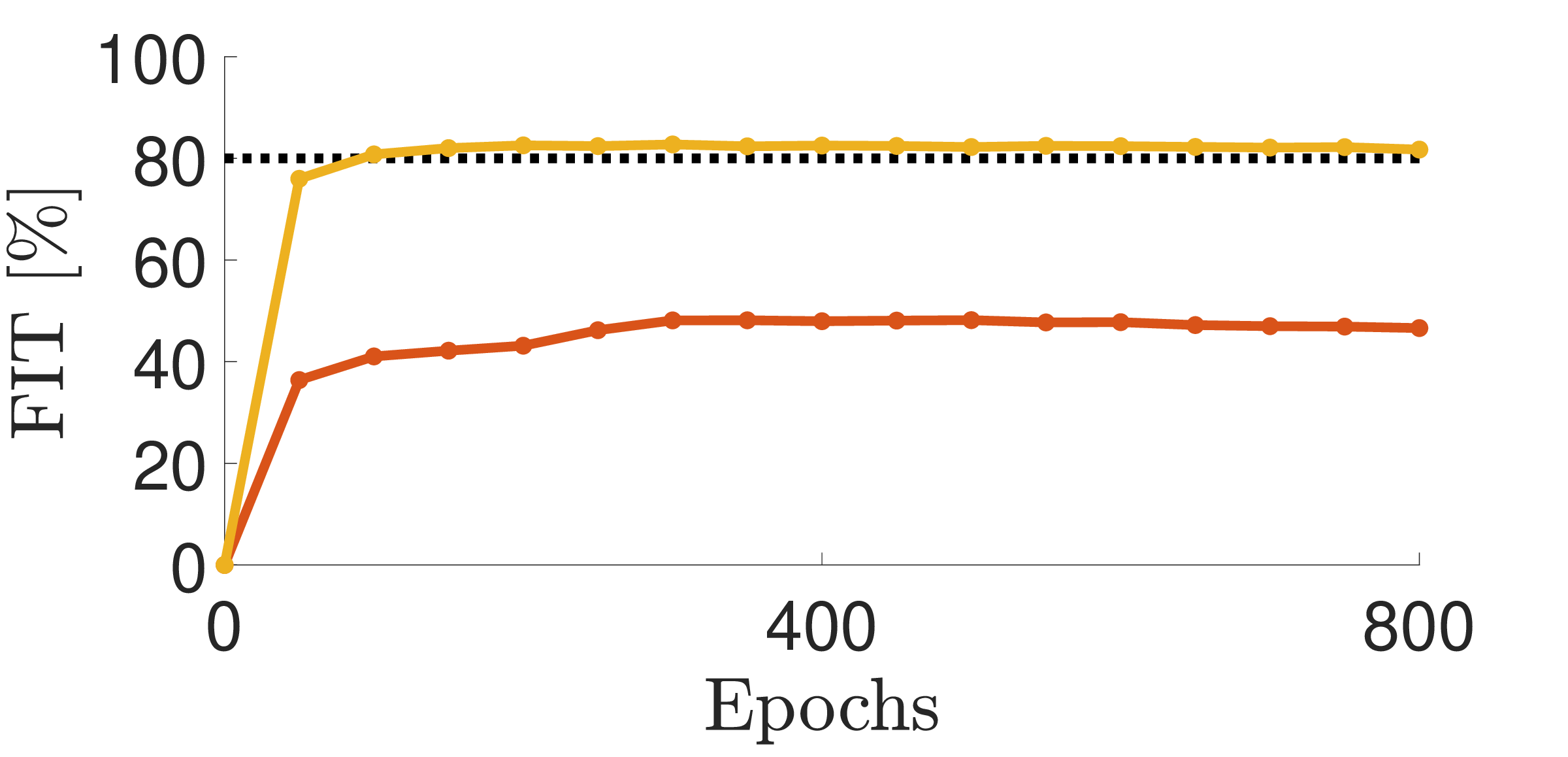} }
	\caption{Sensitivity analysis. The 80\% FIT is represented in dotted black. (a) FIT trend of a 90-state (orange), 54-state (yellow), and 30-state (purple) standard GRU, trained with a 15690-sample dataset; (b) FIT trend of a 90-state (orange), 54-state (yellow), and 30-state (purple) PI-GRU, trained with a 15690-sample dataset; (c) FIT trend of a standard GRU (orange) and of a PI-GRU (yellow), both having 54 states and trained with a 7845-sample dataset. In (b) and (c) only the first 300 and 800 epochs, respectively, are displayed as overfitting occurs thereafter.}
	\label{fig:sensitivityneur}
\end{figure}
\medskip
\subsubsection*{Sensitivity analysis}
\phantom{a}\\[0.15cm]
A short sensitivity analysis is here reported to assess the robustness of the proposed PI-RNN method.

\smallskip
First, the performances of the standard GRU and PI-GRU networks are evaluated using different amount of states, i.e., $n_x=30$ and $n_x=90$. As visible from Figure \ref{fig:sensitivityneur}(a) and \ref{fig:sensitivityneur}(b), the PI-GRU still achieves superior performances, yielding always a FIT above 80\%, which slightly decreases as the state dimension drops. In fact, the PI-GRU states dimension has an impact solely on the number of epochs required to exceed the 80\% FIT. After several tests, the average FIT, $\underline{R}^2$ and $\overline{R}^2$ values, together with their standard deviation, are computed and reported in Table \ref{table:fit90neuroni}. Ultimately, PI-GRUs, unlike standard GRUs, are shown in practice to be robust with respect to the number of neurons.

\smallskip
In a second test, a dataset of 7845 samples (i.e., half of the original dataset) is used to train both the 54-state standard GRU and the PI-GRU one. Once again, the latter outperforms the standard GRU, as evident from Figure \ref{fig:sensitivityneur}(c). Multiple tests are carried out and the obtained average FIT, $\underline{R}^2$ and $\overline{R}^2$ values, together with their standard deviation, are reported in Table \ref{table:piccolodataset}. 
To conclude, another advantage of such physics-informed method lies in the fact that, when only a limited amount of data is available, the PI-GRU, contrarily to the standard GRU, is still capable of modelling the system dynamics. Additionally, the 54-state PI-GRU trained with a reduced dataset (see Table \ref{table:piccolodataset}) performs better even than the 90-state standard GRU trained with the complete dataset (see Table \ref{table:fit90neuroni}).
\smallskip
\begin{remark}
	For a complete analysis, the comparison of a standard GRU and another one characterized by a physics-informed loss function \cite{karpatne2017physics} is carried out. In DHSs, for instance, the load supply temperature is always greater than the output one. However, in the AROMA DHS case, when the GRU is trained by minimizing the standard loss function with the addition of the just mentioned constraint, a performance improvement is not evident with respect to standard approaches. Thus, this physics-informed method has been discarded.
\end{remark}
\begin{table}[t]
	\centering
	\caption{Comparison between the performance of a standard GRU and a PI-GRU, having 90 and 30 states, trained with 15690 samples.}
	\begin{tabular}{c|ccccc}
		\!\!&\!$n_x$ & \!\!\!$n_x^{[i]}$ & \!FIT [\%]\!\!& $\underline{R}^2$ [\%]\!\!& \!\!$\overline{R}^2$ [\%]\!\!\!\!\!\\ [0.1cm] \hline  \\[-6px] 
	\!\!	\!\!GRU\! \!&\!90&\!\!\![15,15,15,15,15,15]\!\! & \!60.4$\pm$6.1\!\! & 42.2$\pm$13.9 \!\!& \!\!97.5$\pm$0.5\!\!\!\!\! \\[2px] 
	\!\!	\!\!PI-GRU\!\! &\!90&\!\!\![9,9,9,16,20,27]\!\! & \!83.7$\pm$0.1\!\! & 87.1$\pm$0.5 \!\! & \! \!99.4$\pm$0.1 \!\!\!\!\!\\[3px] \hdashline\\[-5px]
	\!\!	\!\!GRU \!\!&\!30&\!\!\![5,5,5,5,5,5]\!\! &  \!43.1$\pm$0.9\!\! & 11.8$\pm$3.2 \!\!&\!\!92.2$\pm$1.3\!\!\!\!\! \\[2px] 
	\!\!	\!\!PI-GRU\! \!& \!30&\!\!\![3,3,3,4,8,9]\!\! & \!82.3$\pm$0.2\!\!  &  85.9$\pm$1.6 \!\!&\!\!98.7$\pm$0.2 \!\!\!\!
	\end{tabular}
	\label{table:fit90neuroni}
\end{table}
\begin{table}[t]
	\centering
	\caption{Comparison between the performance of a standard GRU and a PI-GRU, both having 54 states and trained with 7845 samples.}
	\begin{tabular}{c|ccccc}
		&$n_x$& $n_x^{[i]}$ & FIT [\%]& $\underline{R}^2$ [\%]& $\overline{R}^2$ [\%]\\ [0.1cm] \hline  \\[-6px] 
		GRU &54& [9,9,9,9,9,9] & 47.2$\pm$1.4 & 10.7$\pm$2.4 & 95.4$\pm$1.3 \\[2px] 
		PI-GRU &54& [6,6,6,8,12,16] & 82.3$\pm$0.4 &  85.3$\pm$1.9 & 99.2$\pm$0.1
	\end{tabular}
	\label{table:piccolodataset}
\end{table}
\subsection{Control results}
The formulated NMPC regulator is implemented in MATLAB R2023a using the CasADi environment and the Ipopt solver. Moreover, the control tests are carried out on the developed AROMA DHS simulator, implemented in the Modelica environment using \cite{DHN4Control}. 

\begin{figure}[t]
	\centering
	\captionsetup[subfloat]{labelfont=scriptsize,textfont=scriptsize}
	\subfloat[]{ \includegraphics[width=0.45\textwidth]{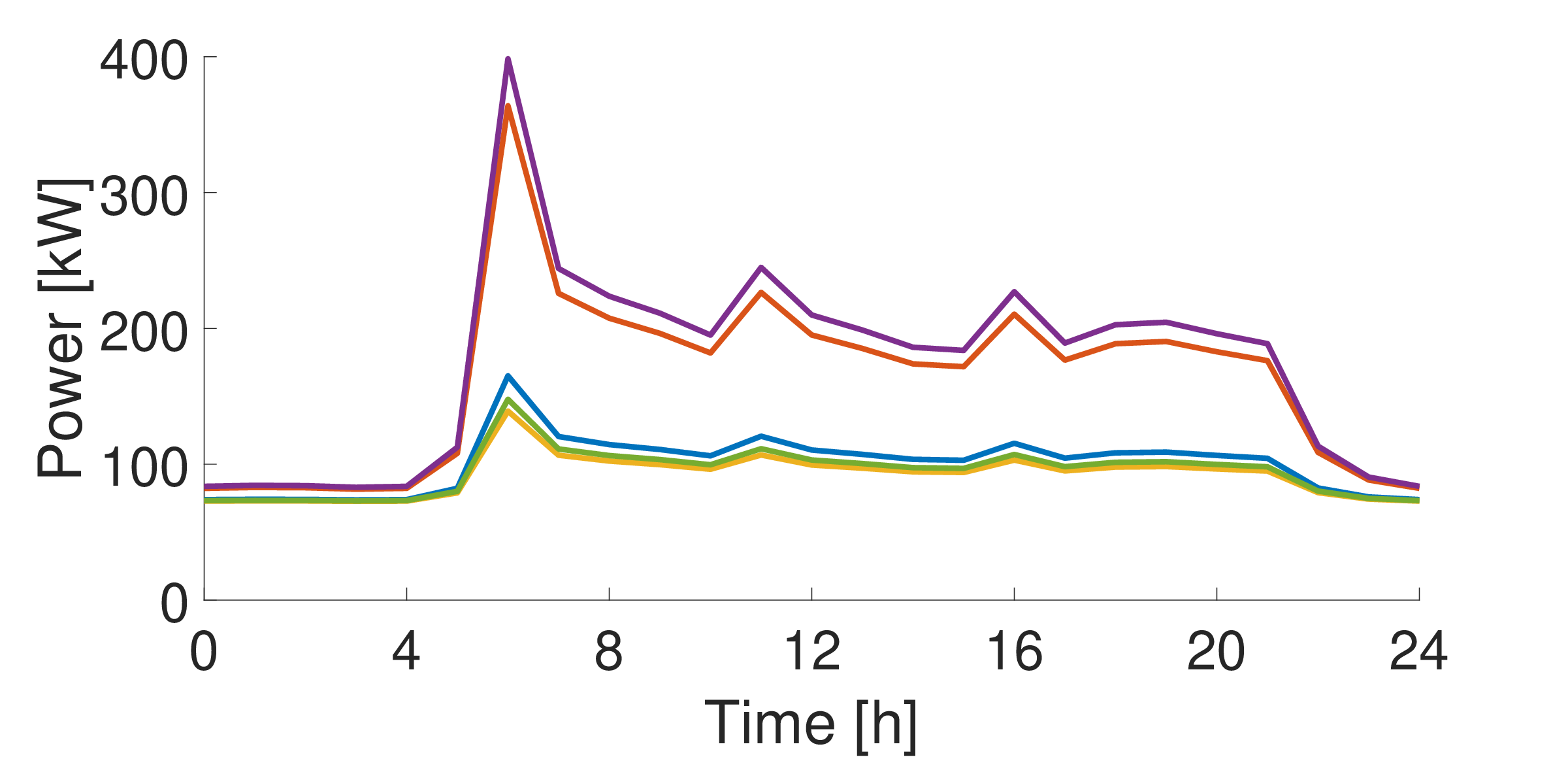} }\hspace{0.2cm}
	\subfloat[]{\includegraphics[width=0.45\textwidth]{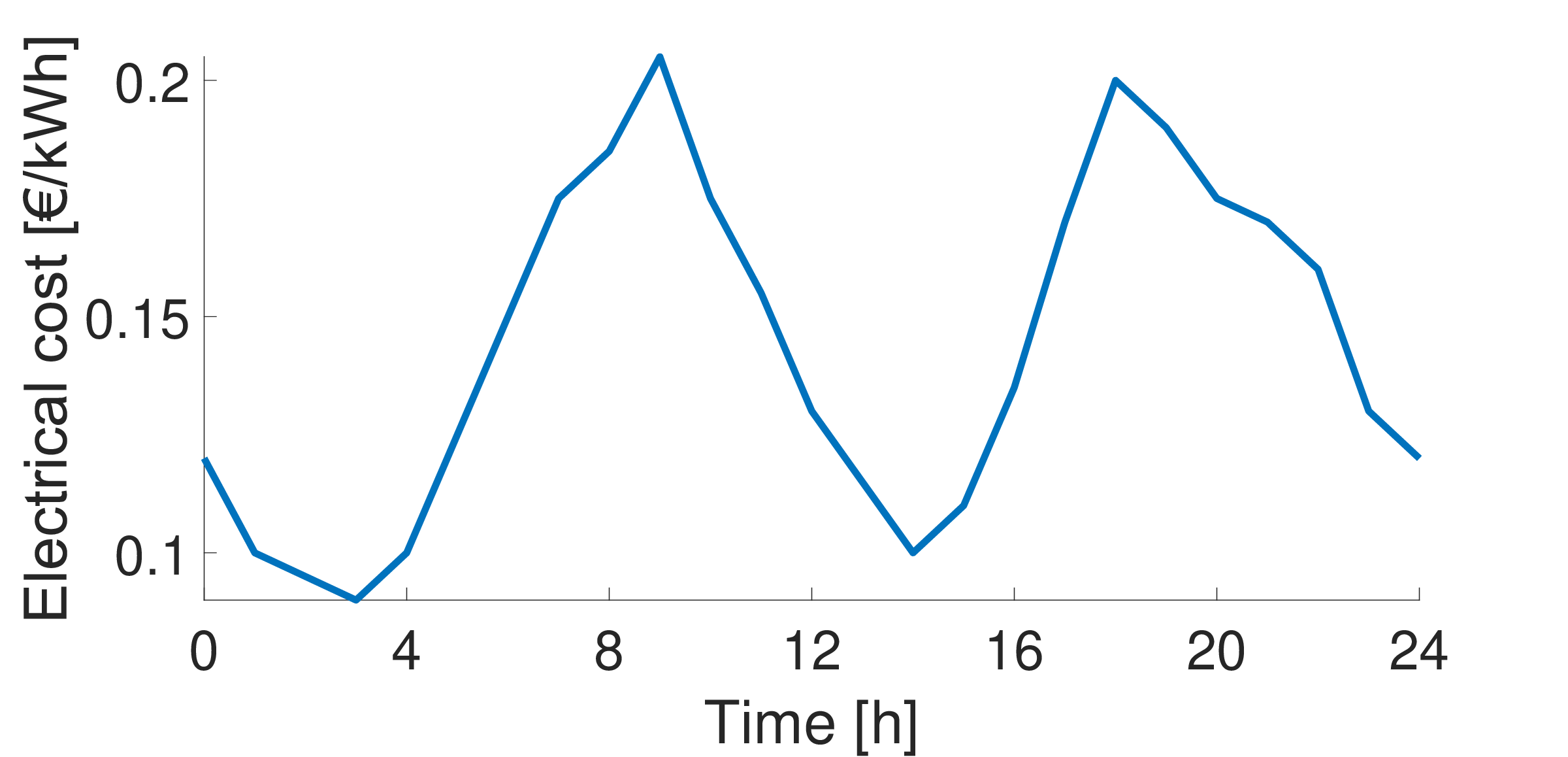} }
	\caption{NMPC inputs. (a) Thermal demands: the first load’s thermal demand is depicted in blue, in purple the second’s, in orange the third’s, in green the fourth’s and in yellow the fifth’s; (b) Daily electrical price $c_{el}$ profile.}
	\label{fig:mpc_inp}
\end{figure}

\smallskip
The NMPC regulator is executed with a sampling time \hbox{$\tau_s = 5$ min} and it considers a prediction horizon {$N = 6 \text{h} / \tau_s = 72$ steps}. The main control design parameters are reported in Table \ref{tab:NMPC_param}. The heating station is assumed to be modelled as an equivalent heat pump, normally characterized by an electrical-to-thermal efficiency $\eta$ larger than one (i.e., the Coefficient of Performance  \cite{blanchard1980coefficient}). The lower bound of the thermal load supply temperatures is time-varying, as previously discussed and reported in Table \ref{tab:NMPC_param}. For the sake of simplicity, an open-loop observer replicating the identified RNN nonlinear dynamics is implemented to provide the state estimate to the NMPC. Moreover, considering a typical DHSs operation, the daily trend of the considered thermal demands $P_i^{\,c}$ is reported in Figure \ref{fig:mpc_inp}(a), whereas the electrical price profile $c_{el}$ is depicted in Figure \ref{fig:mpc_inp}(b).

\begin{table}[b!]
	\caption{NMPC control parameters.} \label{tab:NMPC_param}
	\centering
	\begin{tabular}{c c | c c }
		\hline
		&&& \\[0.00000000001cm]
		$N$ & $72$ & $\tau_s$  & $5$ min \\[0.1cm]
		$N_b$ & $6$ & $\eta$ & $2.5$ \\[0.1cm]		
		$c_t$ & $10$ & $T^{\star}$ & $75$°C \\[0.1cm]
		$\overline{\Delta T}_{\,0}^{\,s}$  & $5$°C & $(\underline{P}_0^{\,s}, \overline{P}_0^{\,s})$ & $(0.1,10)$ MW \\[0.1cm]	
		$(\underline{T}_0^{\,s}, \overline{T}_0^{\,s})$ & $(65,85)$°C &  $(\underline{T}_0^{\,r}, \overline{T}_0^{\,r})$ & $(40,70)$°C  \\
		$\overline{T}_i^{\,s}(k)$ & $85$°C, $\forall k$ & $\underline{T}_i^{\,s}(k)$ &
		\begin{minipage}{3cm} 
			\begin{equation*}		
				\left\{	
				\begin{aligned}			
					& 70^{\circ}\text{C}, \;  84 \leq k \leq 228\\
					& 65^{\circ}\text{C}, \; \text{otherwise}
				\end{aligned} \right.
			\end{equation*}
		\end{minipage} \\[0.5cm]
		\hline
	\end{tabular}
\end{table}

Regarding the model choice, the NMPC performances are tested considering the 54-state standard GRU and the 30-state PI-GRU model. On the one hand, the 54-state standard GRU is selected in order to have at least an average FIT of 50\% (Table \ref{table:fit54neuroni}) and a computationally tractable optimization problem. Indeed, the 90-state standard GRU yields slightly higher FIT values (Table \ref{table:fit90neuroni}) but yet computationally heavy optimization problems in case of multi-step-ahead prediction horizons, because of the model large dimension: the solver takes almost three minutes per iteration, which would introduce unacceptable delays considering $\tau_s=5$ min. By contrast, the 30-state standard GRU is computationally lighter but it leads to very poor identification accuracy (Table \ref{table:fit90neuroni}) and therefore to unreliable models. On the other hand, the 30-state PI-GRU is the physics-informed model characterized by the minimum amount of states that reaches a FIT of 80\% (Table \ref{table:fit90neuroni}), thus producing accurate predictions and also computationally tractable problems. In fact, note that the 30-state PI-GRU model is characterized, by considering input, output and state variables, by $n_\text{v} = 53$ variables, whereas the 54-state standard GRU is characterized by $n_v=77$ variables. By contrast, the AROMA DHS physical model is characterized, overall, by $n_\text{v} = 882$ variables\footnote{The number of variables of the physical model is returned by the DHS simulator implemented in Modelica.}. Therefore, the amount of optimization variables that an NMPC regulator has to manage is $n_\text{v} \cdot N$, where, in case of data-based models is a tractable number, whereas in case of physical models is clearly intractable.

\bigskip
The NMPC exploiting the standard GRU model and the one exploiting the PI-GRU one are also compared with a rule-based strategy, where the heating station is operated at constant supply temperature, as typical in DHSs \cite{la2023optimal,vivian2020load}, by setting $T_0^{\,s} = \text{75}^{\circ}\text{C}$.

\bigskip
The control strategies are tested over a daily simulation and evaluated based on the following performance indexes: the daily production cost \mbox{$C_p=\sum\limits_{t=1}^T c_{el}(t) P_0^*(t)/{\eta}$}, where $P_0^*(t)$ is the effective power produced by the heating station and $T=24$ h$/\tau_s=288$,  the average computational time $t_{\text{avg}}$, and the total thermal losses $P_{\text{loss}}=\sum\limits_{t=1}^T \big (P_0^*(t)-\sum\limits_{i=1}^{n_c} P_i^{\,c}(t) \big)$.

\bigskip
The computed indexes are reported in Table \ref{table:MPCcomp} for the three analysed control strategies.
In particular, when the NMPC regulator exploits the PI-GRU model, the computational time is significantly reduced, given that the 54-state standard GRU is characterized by a larger dimension. Moreover, thanks to the greater reliability of PI-GRU predictions, the production cost is lowered as well. 
By contrast, when the rule-based strategy is adopted, the production cost clearly grows with respect to NMPC strategies. These economic savings are particularly encouraging in terms of efficiency improvement, as thermal losses are significantly reduced when using NMPC strategies, and in particular when exploiting the PI-GRU-based NMPC, as evident from Table \ref{table:MPCcomp}. Moreover, it is worth noting that if cogeneration or thermal storages were present, further savings would be achieved. Ultimately, by using computationally-efficient data-based models, the optimization problem is tractable even with multi-step-ahead prediction horizons: the computational complexity issues related to the physical model are overcome.

\begin{table}[t!]
	\centering
	\caption{Comparison among adopted control methods.}
	\begin{tabular}{c|ccc}
		& $C_p$ & $t_{\text{avg}}$ & $P_{\text{loss}}$ \\ [0.1cm] \hline \\[-6px] 
		{NMPC (30-state PI-GRU)} & 881.4\,\euro & 17\,s & 10.4\,kW  \\[2px] 
		{NMPC (54-state standard GRU)} & 907.2\,\euro
		& 40\,s & 16.0\,kW \\[2px]
		{Rule-based strategy} & 936.8\,\euro
		& - & \!\!32.0\,kW
	\end{tabular}
	\label{table:MPCcomp}
\end{table}
\begin{figure}[t!]
	\centering
	\captionsetup[subfloat]{labelfont=scriptsize,textfont=scriptsize}
	\subfloat[]{ \includegraphics[width=0.45\textwidth]{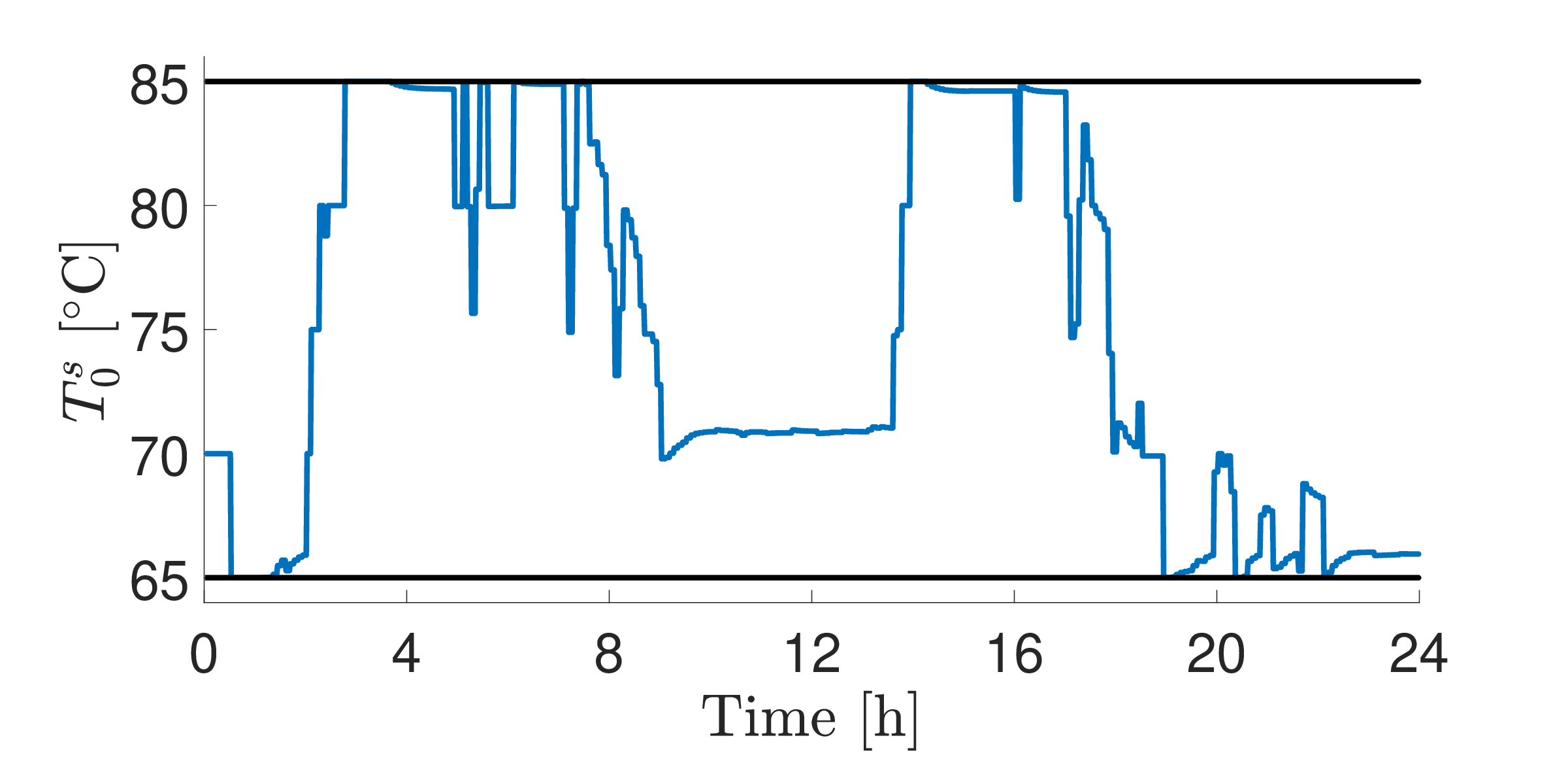} }\hspace{0.2cm}
	\subfloat[]{\includegraphics[width=0.45\textwidth]{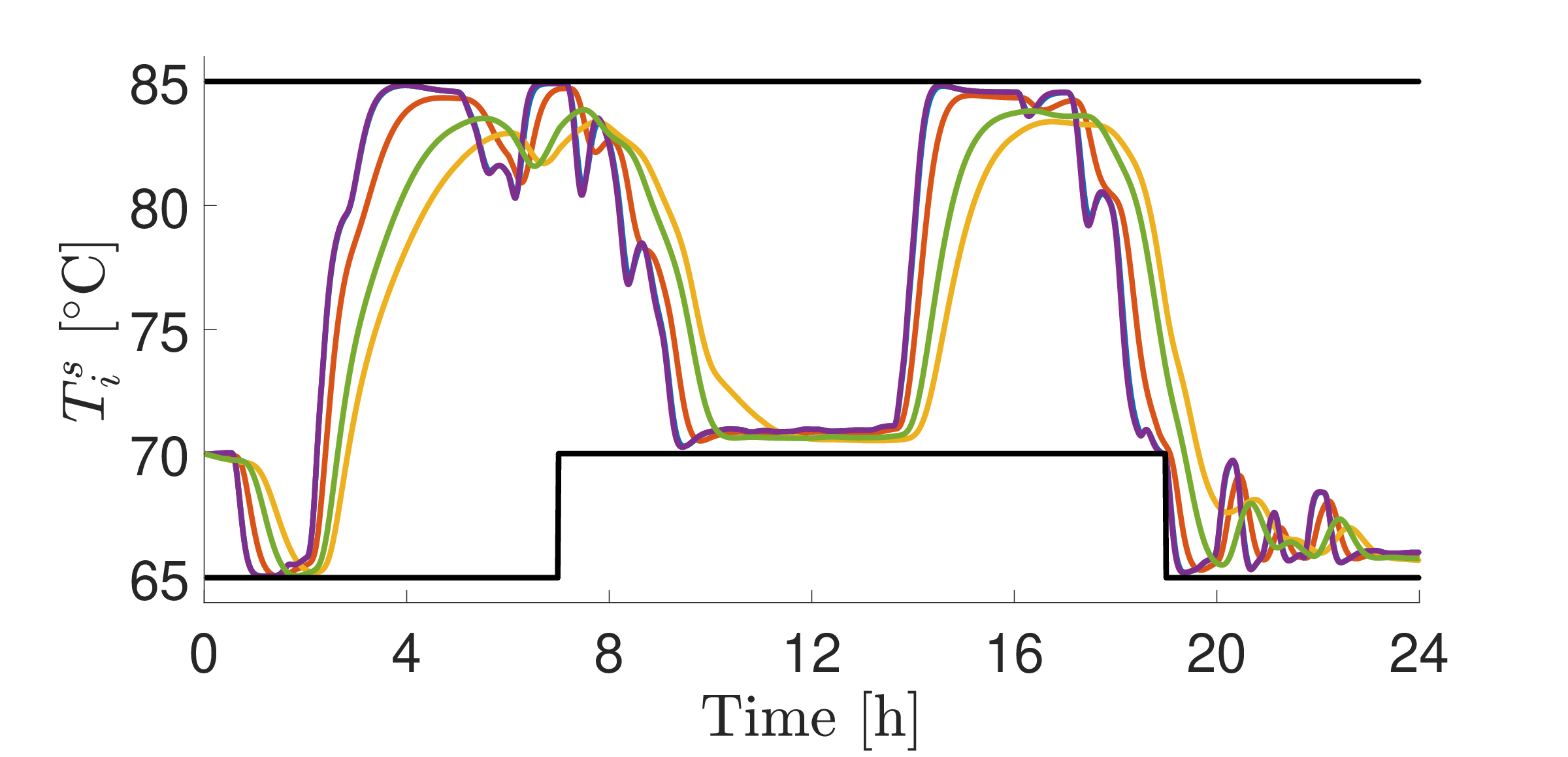} }
	\caption{NMPC results over a daily optimization, for the AROMA DHS. Constraints are depicted in black solid lines. (a) Optimized supply temperature at the heating station; (b) loads supply temperatures: the first load’s supply temperature is represented in blue, in purple the second’s, in orange the third’s, in green the fourth’s and in yellow the fifth’s.}
	\label{fig:mpc}
\end{figure}

\bigskip
The trends of the control variable $T_0^{\,s}$ and of the five load supply temperatures obtained by executing the PI-GRU-based NMPC for a whole day, are reported in Figure \ref{fig:mpc}. As visible from Figure \ref{fig:mpc}(a), the heating station supply temperature never exceeds its bounds. 
Note that when the electrical price reaches its peak, as shown in Figure \ref{fig:mpc_inp}(b), the NMPC decreases the heating station supply temperature. On the other hand, $T_0^{\,s}$ is raised when the electrical price reaches its minimum. 
This predictive ability enables to reduce the production costs, as the DHS network is charged by raising the supply temperature when convenient.
Moreover, given that the PI-GRU embeds the network thermal dynamics, the NMPC is able to optimize the DHS operations while also ensuring that thermal loads are always supplied with water at temperature within prescribed limits, despite their distance from the heating station, as evident from Figure \ref{fig:mpc}(b).

\section{Conclusions}\label{sec:concl}
A novel data-based modelling methodology and optimal control strategy are proposed for District Heating Systems (DHSs). In fact, they typically involve large-scale nonlinear dynamical models, which are not suited for online optimization-based strategies. On the other hand, DHSs are characterized by significant amounts of historical data, which can be leveraged to identify computationally-efficient data-based models, e.g., through the use of Recurrent Neural Networks (RNNs). 
This work firstly proposes a novel modelling approach where the potential of RNNs is combined with a commonly known physical information in DHSs, i.e., the DHS network topology, leading to the design of Physics-Informed RNN (PI-RNN) models. It is shown that interconnecting multiple RNNs by resembling the DHS network topology leads to significant improvements in terms of faster training procedures, higher identification accuracy, and reduced modelling complexity, with respect to pure black-box RNN methods.
The developed PI-RNN model is leveraged for the design of an NMPC regulator, able to minimize production costs, increase system efficiency and respect operative constraints over the whole DHS network. The proposed PI-RNN-based NMPC strategy enables to optimize the DHS with a prediction horizon of few hours and reduced computational times, obtaining enhanced performances with respect to  the standard
RNN-based NMPC and a rule-based control strategy. 
The proposed methods are tested on a DHS benchmark referenced in the literature, i.e., the AROMA DHS, implemented in simulation using the Modelica environment, achieving promising results both from the modelling and control perspective.

The proposed PI-RNNs approach can be actually applied to other types of networked systems where the topology of physical interactions among subsystems is known, such as industrial plants, electrical and gas grids, or biological systems. Thus, future related works regard the development of a methodological approach to design physics-informed data-based models of networked system, which can be easily extended to other applications. Moreover, it is worth investigating lifelong learning algorithms for physics-informed models, able to adapt, through a continuous information acquisition, existing models to exogenous changes, e.g., given by a variation in the system interconnections topology.

\section*{Acknowledgments}
This work of Laura Boca de Giuli and Riccardo Scattolini was carried out within the MICS (Made in Italy – Circular and Sustainable) Extended Partnership and received funding from Next-Generation EU (Italian PNRR – M4 C2, Invest 1.3 – D.D. 1551.11-10-2022, PE00000004). CUP MICS D43C22003120001

\bibliographystyle{IEEEtran}
\bibliography{Bibliography}


\begin{IEEEbiography}
	[{\includegraphics[width=1in,height=1.25 in,clip,keepaspectratio]{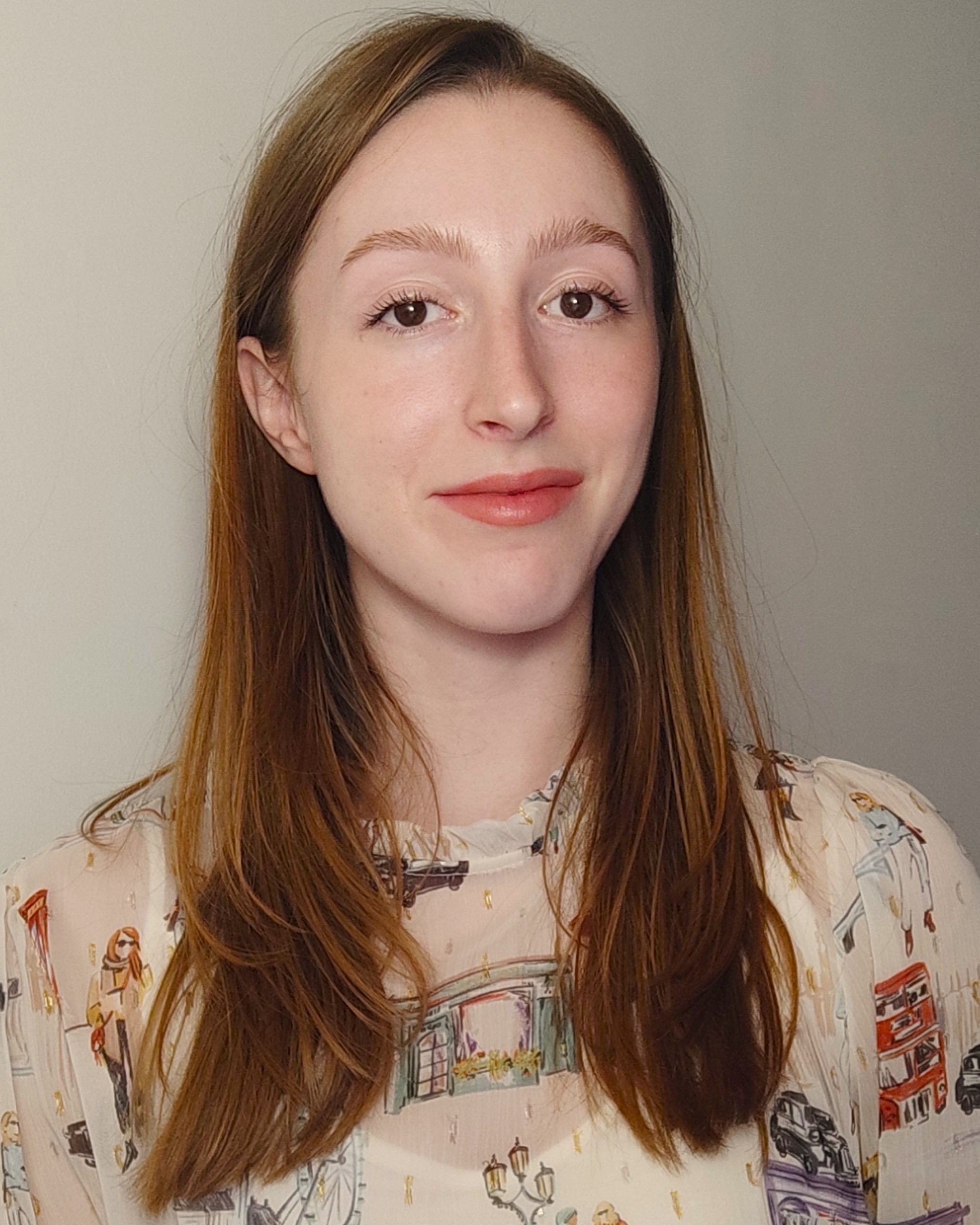}}]{Laura Boca de Giuli} received the B.Sc. and M.Sc. cum laude in Automation and Control Engineering from Politecnico di Milano in September 2020 and in May 2023, respectively. Her Master’s thesis concerned modelling, predictive control and lifelong learning applied to district heating systems through physics-informed neural network techniques. In 2021, she spent a semester as exchange student at the École Polytechnique Fédérale de Lausanne (Switzerland). In June 2023, she enrolled in the PhD programme of Information Technology - Systems and Control area - at Politecnico di Milano. Her research interests concern the modelling, control and supervision of industrial and energy plants via machine learning methods.
\end{IEEEbiography}

\begin{IEEEbiography}
	[{\includegraphics[width=1in,height=1.25 in,clip,keepaspectratio]{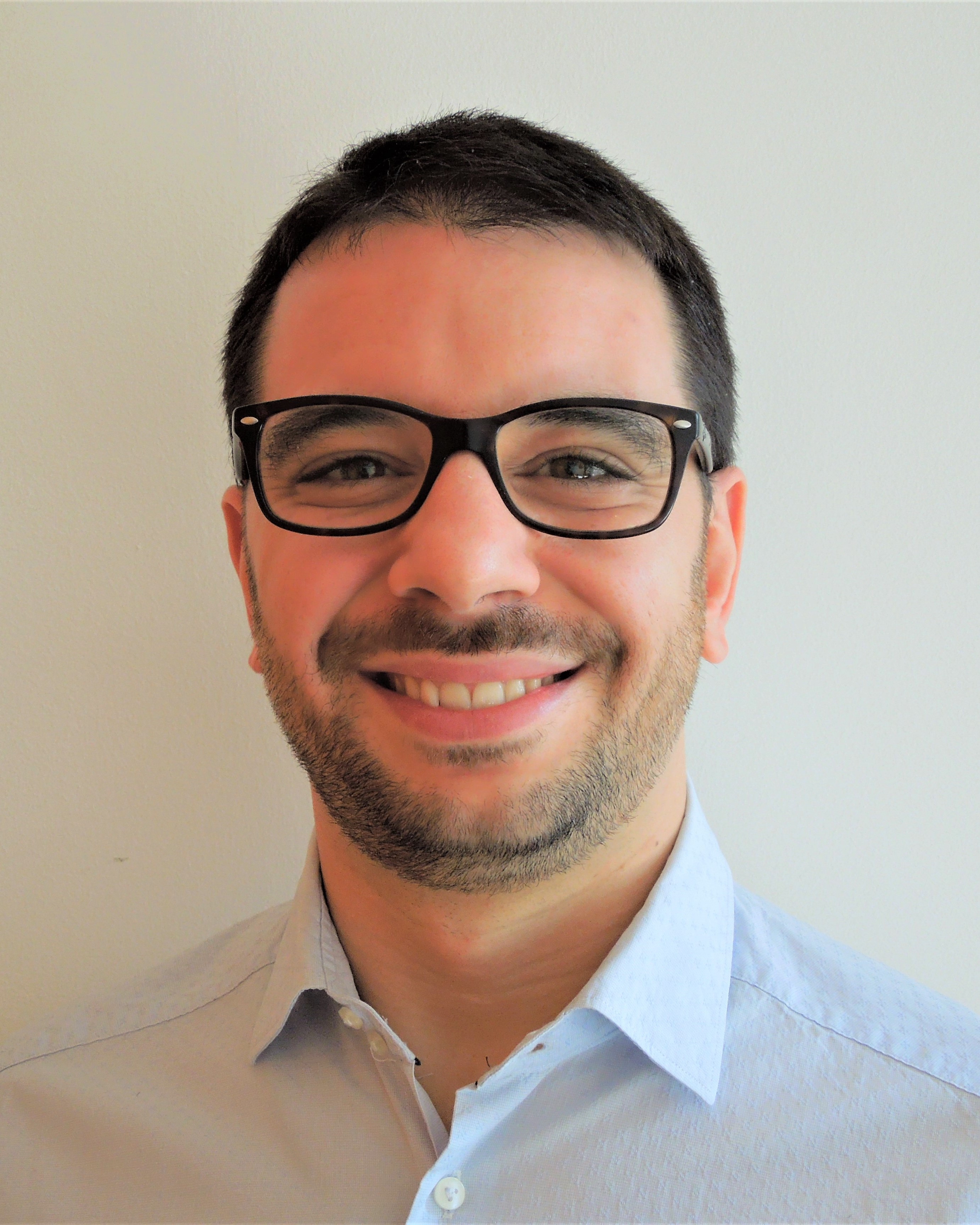}}]{Alessio La Bella} received the B.Sc. and M.Sc. in Automation Engineering cum laude from Politecnico di Milano in 2013 and 2015, respectively. In 2016, he received the Alta Scuola Politecnica Diploma, together with the M.Sc. in Mechatronics Engineering cum laude at Politecnico di Torino. He received the Ph.D. Degree cum laude in Information Technology at Politecnico di Milano in 2020. In 2018, he was visiting researcher at the Automatic Control Lab of the École Polytechnique Fédérale de Lausanne, Switzerland.
	From 2020 to 2022, he worked as Research Engineer at Ricerca sul Sistema Energetico - RSE SpA, designing and implementing advanced predictive control systems for district heating networks and large-scale battery plants, in collaboration with industrial companies and energy utilities. In 2022, he joined Politecnico di Milano as Assistant Professor at Dipartimento di Elettronica, Informazione e Bioingegneria. He is Associate Editor for the International Journal of Adaptive Control and Signal Processing and for EUCA Conference Editorial Board. His research interests concern the theory and design of predictive, multi-agent and learning-based control systems, with particular emphasis on practical challenges arising from the upcoming energy transition. He was recipient of the Dimitris N. Chorafas Prize in 2020.
\end{IEEEbiography}

\begin{IEEEbiography}
	[{\includegraphics[width=1in,height=1.25 in,clip,keepaspectratio]{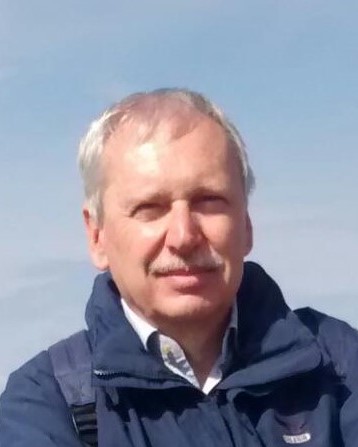}}]{Riccardo Scattolini} is Full Professor of Automatic Control at Politecnico di Milano, Italy. He was awarded Heaviside Premium of the Institution of Electrical Engineers, United Kingdom, and was Associate Editor of the IFAC journal Automatica. His main research interests include modelling, identification, simulation, diagnosis, and control of industrial plants and energy systems, with emphasis on theory and applications of Model Predictive Control and fault detection methods to large-scale and networked systems.
\end{IEEEbiography}

\end{document}